\documentclass[preprint]{aastex}
\usepackage{natbib}
\usepackage{mathrsfs}
\usepackage{rotating}
\usepackage{graphicx}

\newcommand{\etal}{et~al.}
\newcommand{\ud}{\mathrm{d}}

\newcommand{\fn}[1]{\footnote{\scriptsize{#1}}}
\newcommand{\rres}{r_{L}}
\newcommand{\Omegares}{\Omega_{L}}
\newcommand{\kappares}{\kappa_{L}}
\newcommand{\Pone}{\citetalias{PorcoSOIScans07}}
\newcommand{\Pthree}{\citetalias{jemodelshort}}
\newcommand{\Eqn}[1]{Eq{#1}.}  
\newcommand{\Fig}[1]{Fig{#1}.}  
\newcommand{\Cassit}{\textit{Cassini}}  
\newcommand{\Voyit}{\textit{Voyager}}  
\defcitealias{PorcoSOIScans07}{Paper~I}
\defcitealias{jemodelshort}{Paper~III}
\slugcomment{To appear in Icarus}

\shorttitle{Radial Structure in Saturn's Rings}
\shortauthors{Tiscareno \etal}

\begin{document}

\title{Cassini Imaging of Saturn's Rings II:\\A Wavelet Technique for Analysis of Density Waves\\and Other Radial Structure in the Rings}
\author{Matthew~S.~Tiscareno$^1$\footnote{Corresponding author:  \tt{matthewt@astro.cornell.edu}}, Joseph~A.~Burns$^{1,2}$, Philip~D.~Nicholson$^1$,\\Matthew~M.~Hedman$^1$, Carolyn~C.~Porco$^3$}
\affil{$^1$ Department of Astronomy, Cornell University, Ithaca, NY 14853\\$^2$ Department of Theoretical and Applied Mechanics, Cornell University, Ithaca, NY 14853\\$^3$ CICLOPS, Space Science Institute, 4750 Walnut Street, Boulder, CO 80301}

\begin{abstract}

We describe a powerful signal processing method, the continuous wavelet transform, and use it to analyze radial structure in \Cassit{} ISS images of Saturn's rings.  Wavelet analysis locally separates signal components in frequency space, causing many structures to become evident that are difficult to observe with the naked eye.  Density waves, generated at resonances with saturnian satellites orbiting outside (or within) the rings, are particularly amenable to such analysis.  We identify a number of previously unobserved weak waves, and demonstrate the wavelet transform's ability to isolate multiple waves superimposed on top of one another.  We also present two wave-like structures that we are unable to conclusively identify.  

In a multi-step semi-automated process, we recover four parameters from clearly observed weak spiral density waves:  the local ring surface density, the local ring viscosity, the precise resonance location (useful for pointing images, and potentially for refining saturnian astrometry), and the wave amplitude (potentially providing new constraints upon the masses of the perturbing moons).  Our derived surface densities have less scatter than previous measurements that were derived from stronger non-linear waves, and suggest a gentle linear increase in surface density from the inner to the mid-A~Ring.  We show that ring viscosity consistently increases from the Cassini Division outward to the Encke Gap.  Meaningful upper limits on ring thickness can be placed on the Cassini Division (3.0~m at $r \sim 118,800$~km, 4.5~m at $r \sim 120,700$~km) and the inner A Ring (10~to 15~m for $r < 127,000$~km).  

\end{abstract}

\textit{Subject headings:}  Saturn, Rings; Resonances, Rings\\
\indent{}\textit{Running header:}  Wavelet Analysis of Density Waves in Saturn's Rings

\section{Introduction}

The radial structure of Saturn's rings has long been the subject of inquiry by astronomers, dating from G.~D.~Cassini's discovery in 1675 of the division that bears his name, and W.~Herschel's realization in 1791 that the Cassini Division appears identical on both the northern and southern ring faces, definitively identifying it as a gap \citep{Alexander62,VanHelden84}.  Many additional ringlets and gaps came to light during succeeding centuries, but more subtle variations of density within a ring awaited discovery by spacecraft.  

In preparation for spacecraft encounters, an analogy with galactic dynamics -- particularly the Lindblad resonances and spiral density waves that cause galaxies to have spiral arms \citep{LS64} -- was first applied to Saturn's rings by \citet{GT78a,GT78b,GT80}, who noted that radial structure could arise due to the perturbations of exterior satellites.  The general theory of resonant interactions within planetary rings was reviewed by \citet{GT82} and by \citet{Shu84}.  

The \Voyit{} encounters with Saturn in 1980 and 1981 produced an unprecedented data set on the rings' radial structure, including evidence for spiral density waves \citep{Cuzzi81,HFL82,Holberg82}.  A number of authors quantitatively analyzed the radial traces of density waves; most of these focused on deriving the ring's local background surface density, which is directly obtained from the wavelength dispersion of density waves.  \citet{Espo83} carried out the first systematic survey for wave traces, firmly establishing the connection with the known locations of Lindblad resonances \citep{LC82}.  \citeauthor{Espo83} additionally attempted to fit the waves' damping length and thus derive ring viscosity.  The analysis of wave traces by \citet{NCP90} concentrated on pinpointing the locations of resonant wave generation, aiding their purpose of precisely navigating the occultation scans and determining Saturn's pole orientation.  \citet{Rosen91a,Rosen91b} wrote extensively on methods of density wave analysis, and additionally fit wave amplitudes in order to derive masses for the perturbing moons.  Finally, \citet{Spilker04} undertook a comprehensive survey of spiral density waves in the \Voyit{} data set.  

A distinct form of radial structure in the rings are the wakes raised by the gravity of a passing nearby moon (here called ``moonlet wakes'' to distinguish them from the ``self-gravity wakes'' mentioned briefly in Section~\ref{ViscosityResults}).  These were described and explained in detail by \citet{Show86}, who used that understanding to pinpoint and then observe the moon Pan embedded in the Encke Gap \citep{Show91}.  Further analysis of Pan wakes by \citet{Horn96} was among the first to use a form of localized Fourier analysis to examine the rings' radial structure \citep[though see also][]{Espo83}.  The same technique was applied to the unexplained radial structure in the B~Ring \citep{HC96} and to spiral density waves in the A~Ring \citep{Spilker04}.  

The Imaging Science Subsystem (ISS) \citep{PorcoSSR04}, on board the \Cassit{} spacecraft, has vastly improved upon the clarity of \Voyit{} images of the rings.  \Cassit{}'s highest-quality imaging to date of Saturn's rings was accomplished during the spacecraft's insertion into Saturn orbit (SOI) on 1~July~2004.  The companion paper by \citet[hereafter \Pone{}]{PorcoSOIScans07} describes the calibration, image processing, and presentation of these data, as well as of images collected during a lower-resolution but contiguous radial sequence across the sunlit side of the rings on 20~May and 21~May~2005, resulting in a series of scans of brightness with orbital radius.  Also in \Pone{} is a catalog of the strongest gravitational resonances and resonance strengths arising from the known moons, including Pan and Atlas, both within and exterior to the rings, derived using the latest \Cassit{} values for moon masses and orbital elements \citep{Jake05,Jake06,Spitale06}

In this paper, we analyze the 1~July~2004 radial scans using techniques involving the wavelet transform \citep{Daub92,Farge92,TC98,Addison02}, which functions as a method of localized Fourier analysis that is optimized to simultaneously detect signals over a wide range of frequencies.  Developing the wavelet techniques first presented by \citet{DPS04}, we employ a multi-step semi-automated process to measure wave parameters that can be inverted to yield ring and moon properties.  In this task, we focus on density waves that are too weak to have been detected by \Voyit{}, not only to avoid duplication of previous work but also because these weaker waves maintain strictly linear dispersion \citep{Shu83,Shu85}, causing our results to be less susceptible to systematic errors.  Additionally, we present several examples of wavelet transforms of the SOI scans, both to further illuminate the structure of Saturn's rings and to demonstrate the advantages of wavelet analysis.  

A brief paper describing our model of the complex spiral density waves generated at resonances with the co-orbital satellites Janus and Epimetheus \citep{jemodelshort} is \Pthree{} in this sequence.  

Section~\ref{DW} reviews the theory of spiral density waves, Section~\ref{WaveletTheory} that of wavelet transforms, and Section~\ref{Synthetic} describes a new wavelet method for analyzing density waves.  Section~\ref{Observations} shows examples of \Cassit{}~ISS radial scans of the rings, with features explicated by wavelet analysis.  Section~\ref{Results} presents the results of our analysis of weak linear density waves.  Section~\ref{Conclusions} presents our conclusions.  

\section{Density Waves \label{DW}}

We begin by recalling the mathematical properties of spiral density waves.  Such waves occur at locations where ring particles are in an inner or outer Lindblad resonance (ILR or OLR) with a perturbing moon \citep{GT82,Shu84}.  At such locations, the frequency of the moon's perturbation forms a simple ratio with the radial (epicyclic) frequency of ring particle orbits, amplifying their eccentricity.  The coherent excitation of ring particles gives rise to compression and rarefaction, initiating a wave that propagates through the disk.  

The density wave generated in a ring at an inner Lindblad resonance (ILR) can be described as a density variation $\Delta \sigma$ on the background surface density $\sigma_0$.  At orbital radius $r$ greater than the resonance location $\rres$, we have \citep{GT82,Shu84,NCP90,Rosen91a}\fn{There are differences in notation among these references.  Firstly, \citet{Shu84} and \citet[hereafter NCP]{NCP90} define $\xi$ to be a factor of $\sqrt{2}$ smaller than our \Eqn{}~\ref{Def_Xi} (which follows \citet{Rosen91a}).  Secondly, all three papers use different normalizations and absorb them into the amplitude $A_L$.  Specifically, $$A_L^{Rosen} =  - \sqrt{\frac{\mathscr{D}_L}{G^3 \sigma_0 r_L}} \frac{A_L^{Shu}}{4\pi^2} =  - \pi^{-1/2} \sigma_0 A_L^{NCP}. $$}
\begin{equation}
\label{DWEq}
\Delta \sigma(r) = \mathrm{Re} \left\{ iA_L e^{-i \phi_0 } 
\left[ 1 - i \xi e^{-i \xi^2/2} \int_{-\infty}^{\xi} e^{i \eta^2/2} \ud \eta \right]  \right\} e^{-( \xi / \xi_D )^3} , 
\end{equation}
\noindent where the dimensionless radial parameter is
\begin{equation}
\label{Def_Xi}
\xi = \left( \frac{\mathscr{D}_L \rres}{2 \pi G \sigma_0} \right)^{1/2} \cdot \frac{r-\rres}{\rres} ,  
\end{equation}
\noindent and further terms are defined below.  

Assuming Saturn's gravity is well described as a point mass plus a $J_2$ harmonic, the factor $\mathscr{D}_L$ is given by \citep{Cuzzi84,MP93}
\begin{equation}
\label{Scripty_D}
\mathscr{D}_L = 3 (m-1) \Omegares^2 + J_2 \left( \frac{r_S}{\rres} \right)^2 \left[ \frac{21}{2} - \frac{9}{2}(m-1) \right] \Omegares^2 , 
\end{equation}
\noindent where the second term is a small correction except for $m=1$.  

The Lindblad resonance occurs at $\rres$.  The local mean motion is $\Omegares$, which must be calculated accounting for the higher-order moments of Saturn's gravity field \citep{MD99}.  Saturn's radius $r_S = \textrm{60,330 km}$ by convention \citep{Kliore80}.  The resonance's azimuthal parameter is $m$, a positive integer equal to the number of spiral arms.  The amplitude $A_L$ is related to the mass of the perturbing satellite, while the damping constant $\xi_D$ describes the ring's viscous response (see Section~\ref{Viscosity}).

\subsection{Initial Phase}

For a given longitude on the rings, $\lambda$, the density wave's initial phase is given by
\begin{equation}
\label{DWInitialPhase}
\phi_0 = m \lambda - (m+k)\lambda_s + k\varpi_s , 
\end{equation}
\noindent where $\lambda_s$ and $\varpi_s$ are the mean longitude and longitude of periapse of the perturbing satellite.  The integers $m$ and $k$ describe the resonance --- a ($k$+1)th-order Lindblad resonance is commonly labeled ($m$+$k$):($m$-1).  A third integer $p$ describes resonances linked to the perturbing satellite's inclination, but no such resonances are discussed in this paper.  

\subsection{Local Wavenumber and Phase \label{WavenumberPhase}}

The integral in \Eqn{}~\ref{DWEq} is a Fresnel integral, which significantly modulates the result near the wave's generation point, but oscillates about unity for higher values of $\xi$.  Downstream, then, the dominant component of \Eqn{}~\ref{DWEq} has the form of a sinusoid with constantly decreasing wavelength (as well as modulating amplitude), such that the wavenumber, $k = 2 \pi / \lambda$, increases linearly with distance from $\rres$ \citep{Shu84}:
\begin{equation}
\label{DWWavenum}
k_{DW}(r) = \frac{\mathscr{D}_L}{2 \pi G \sigma_0 \rres} (r-\rres) . 
\end{equation}

Since the wavenumber increases linearly with $r$, the accumulating phase naturally increases quadratically.  Indeed, making use of $\int k \ud r = \xi^2 / 2$, the instantaneous phase of the sinusoid far from $\xi=0$ (i.~e., $r=\rres$), as a function of radial location $\xi(r)$, asymptotically approaches  \citep{Shu84}
\begin{equation}
\label{DWPhase}
\phi_{DW} = \phi_0 + \xi^2/2 + \pi/4 .
\end{equation}
\noindent It is important to note the term of $\pi/4$ (or 45$^\circ$) added to this quantity.  This asymptotic limit is a good approximation only for $\xi \gtrsim 1$ \citep{NCP90}.

\subsection{Ring Viscosity and Vertical Scale Height \label{Viscosity}}

When the damping constant $\xi_D \gg 1$ (that is, when damping is inefficient enough to allow at least several wavecycles), the local ring kinematic viscosity $\nu$ can be estimated as \citep{GT78b,Shu84}
\begin{equation}
\label{DWViscosity} 
\nu = \frac{9}{7 \kappares \xi_D^3} \left( \frac{\rres}{\mathscr{D}_L} \right)^{1/2} \left(2 \pi G \sigma_0 \right)^{3/2} , 
\end{equation}
\noindent where, for this purpose, the radial frequency $\kappares$ is approximately equal to the orbital frequency $\Omegares$.  

The viscosity $\nu$ is directly related to the rms random velocity $c$ \citep{GT78a}
\begin{equation}
\label{ViscosityRMSVelocity}
c^2 = 2 \nu \Omega \left( \frac{1+\tau^2}{\tau} \right) ,  
\end{equation}
\noindent where $\tau$ is the local optical depth.  \Eqn{}~\ref{ViscosityRMSVelocity} assumes that ring particle interactions are isolated two-particle collisions; it is not valid when the ring particle density is high enough that particle size becomes important \citep{WT88}.  This assumption is probably valid for the A~Ring. 

Under the assumption that random velocities are isotropic, the ring's vertical scale height can be estimated as $H \sim c/\Omega$.  However, this assumption is violated in much of the A~Ring (Section~\ref{ViscosityResults}).  

\subsection{Summary}
The density wave equation (\Eqn{}~\ref{DWEq}) contains five parameters that may vary from wave to wave (not counting the azimuthal parameter $m$, which is fixed by the identity of the resonance), and which are sensitive to physical quantities concerning the ring or the perturbing moon.  
\begin{enumerate}
\item The background surface density of the ring, $\sigma_0$, governs the rate at which the wavenumber increases with distance from resonance.  
\item The resonance location, $\rres$, fixes the wave against translation in the radial direction.  
\item The initial phase, $\phi_0$, is fixed by the moon's orbital phase relative to the location at which the wave is observed. 
\item The damping parameter, $\xi_D$, governs the location at which the wave's amplitude ceases to grow and begins to decay; it is sensitive to the dynamical viscosity of the ring.  
\item The amplitude, $A_L$, governs the overall strength of the perturbation (irrespective of its shape); it is sensitive to the mass of the perturbing moon. 
\end{enumerate}

\section{Wavelet Theory \label{WaveletTheory}}

This paper presents the first thorough analysis of periodicities in the radial structure of Saturn's rings using the wavelet transform, a powerful method of signal processing that has been applied successfully in many other fields.  Since this method is complex, we review its properties in this section, emphasizing aspects that are relevant to our purposes.  For more detailed information on the wavelet transform and its applications, see \citet{Addison02} and references therein.  

\subsection{The Morlet Wavelet \label{TheMorletWavelet}}

The \textit{mother wavelet} governs the way in which the wavelet transform transfers spatial information into the frequency domain.  The mother wavelet we use is the Morlet wavelet, which is optimally suited for identifying oscillatory components of a signal.  

A wavelet transform using the Morlet mother wavelet can be thought of as a localized Fourier transform.  As with standard Fourier analysis, components of the input signal are separated in the frequency domain, but wavelet analysis additionally pinpoints the locations in the signal at which a given frequency is important.  This is particularly useful when the signal's dominant frequency changes with location (as it does in a density wave propagating through a disk).  The advantage of wavelet analysis over other time-frequency methods (such as the windowed Fourier transform) lies in the automatic scaling of the window along with the interrogating waveform, yielding optimal simultaneous detection of both high- and low-frequency signal components.  

The Morlet wavelet is simply a complex sinusoid within a Gaussian envelope:
\begin{equation}
\label{morlet}
\psi(t) = \pi^{-1/4} \exp \left[ i \omega_0 t - t^2/2 \right] .  
\end{equation}
\noindent The wavelet's central frequency, $\omega_0$, is selected beforehand by the user (see Section~\ref{ChoosingOmega}).  

\begin{figure}[!t]
\begin{center}
\includegraphics[width=10cm,keepaspectratio=true]{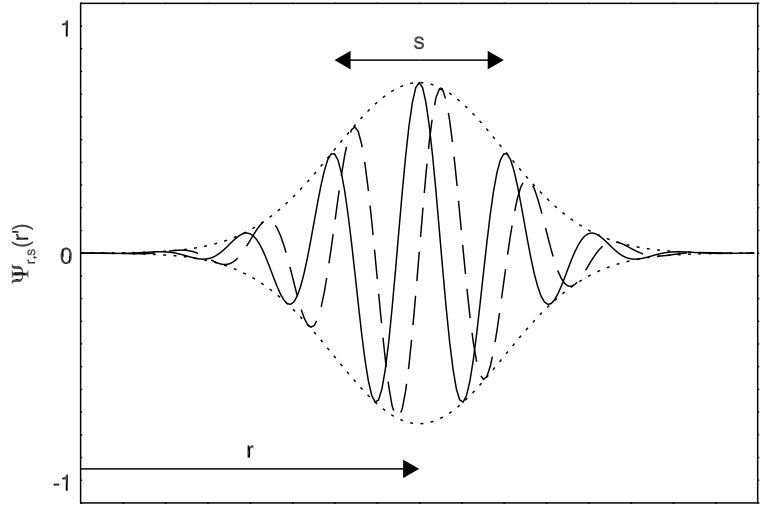}
\caption{The Morlet wavelet with $\omega_0=6$.  The solid line shows the real part, the dashed line the imaginary part.  For ease of viewing, the gaussian envelope within which the wavelet oscillates is shown as a dotted line.  The wavelet translates and dilates in order to interrogate the signal at all scales ($s$) and locations ($r$). \label{morletfig}}
\end{center}
\end{figure}

Wavelet analysis is carried out by translating and dilating the mother wavelet (see \Fig{}~\ref{morletfig}), and convolving these \textit{daughter wavelets} with the input signal (see \Eqn{s}~\ref{WT} and~\ref{WTFT}).  At a given location $r$ and spatial scale $s$, the daughter wavelet is given by \citep{Addison02}
\begin{equation}
\label{morlet1}
\psi_{r,s}(r^\prime) = s^{-1/2} \pi^{-1/4} \exp \left[ \frac{i \omega_0 (r^\prime-r)}{s} - \frac{(r^\prime-r)^2}{2s^2} \right],  
\end{equation}
and the daughter of the Fourier-transformed mother wavelet is given by
\begin{equation}
\label{morlet2}
\hat{\psi}_{r,s}(\omega) = (2s)^{1/2} \pi^{1/4} \exp\left[-i \omega r - \frac{(s\omega-\omega_0)^2}{2} \right] . 
\end{equation}

\subsection{The Wavelet Transform \label{waveletintro}}

For an evenly sampled radial signal, $x(r)$, the continuous wavelet transform (CWT)\fn{We use a discretized form of the CWT, given the digital nature of our signal.  This is not to be confused with the discrete wavelet transform (DWT), which is based on fundamentally different mathematics \citep{Daub92,Graps95,Mallat98}.  For analysis of a feature in Saturn's rings using DWT methods, see \citet{Bendjoya93} and \citet{Spahn93}.} is given by \citep{TC98,Addison02}
\begin{equation}
\label{WT}
T(r,s) = \int^{\infty}_{-\infty} x(r^\prime) \psi^{*}_{r,s}(r^\prime) \ud r^\prime ,  
\end{equation}
\noindent where $^{*}$ denotes the complex conjugate.  In practice, the wavelet transform is more efficiently calculated in terms of the product of the Fourier transforms,
\begin{equation}
\label{WTFT}
T(r,s) = \int^{\infty}_{-\infty} \hat{x}(\omega) \hat{\psi}^{*}_{r,s}(\omega) \ud \omega .  
\end{equation}

For more discussion, see \citet{Farge92} and \citet{TC98}, the latter of whose software\fn{Software available at http://atoc.colorado.edu/research/wavelets/} we use for this purpose.  

\subsection{Choosing the Central Frequency \label{ChoosingOmega}}

Wavelet analysis follows a form of the Heisenberg uncertainty principle, in that an increase in spatial resolution comes only at the cost of decreased frequency resolution, and vice versa \citep[see][]{Addison02}.  Classical Fourier analysis can be seen as an end-member in this trade-off, providing the optimum possible frequency resolution, but at the cost of having no spatial resolution at all.  For the Morlet wavelet, this trade-off is governed by the central frequency $\omega_0$ in \Eqn{s}~\ref{morlet} to~\ref{morlet2}.  Higher values of $\omega_0$ cause the Gaussian envelope to contain more oscillations of the sinusoid, and consequently yield higher spectral resolution at the expense of lower spatial resolution (see \Fig{}~\ref{PanWakes} for an example).  In this paper, unless otherwise specified, we use a value of $\omega_0 = 6$ (although there is nothing intrinsically special about any particular value of $\omega_0$, which need not even be an integer).  A typical Morlet wavelet can be seen in \Fig{}~\ref{morletfig}.  

We should note that the expression for the Morlet wavelet given in \Eqn{s}~\ref{morlet} to~\ref{morlet2} is only valid for $\omega_0 \gtrsim 5$.  For smaller central frequencies, one cannot ignore the addition of a real-valued constant to the sinusoid, which is necessary to ensure compliance with the requirement that an admissible wavelet must have zero mean \citep{Farge92,Addison02}.

\subsection{Wavelength and Wavenumber}

The scale parameter, $s$, is directly proportional to the more familiar Fourier wavelength, $\lambda_F$.  For the Morlet wavelet, this relationship is given by \citep{TC98}
\begin{equation}
\label{FourierScale}
\lambda_F = \frac{4 \pi s}{\omega_0 + \sqrt{2+\omega_0}}  .
\end{equation}
\noindent  The wavelet plots in this paper are shown with a $y$-axis that increases linearly with the wavenumber $k = 2 \pi / \lambda_F$ (and thus inversely with $s$), as the trace of a density wave appears linear in that representation (see \Eqn{}~\ref{DWWavenum}).  

\subsection{Inverse Wavelet Transform}

The wavelet transform can be inverted, recovering the input signal to good accuracy.  Before inverting, a powerful method of filtering can be implemented by setting undesired elements in the transform array to zero.  The inverse wavelet transform can be written \citep{Farge92}
\begin{equation}
\label{InverseWavelet}
x(r) = \frac{1}{C_\delta} \int^{\infty}_{0} \frac{T(r,s)}{s^{3/2}} \ud s .  
\end{equation}
\noindent The normalization constant is 
\begin{equation}
C_\delta = \int^{\infty}_0 \frac{\hat{\psi}_{0,1}(\omega)}{\omega} \ud \omega , 
\end{equation}
\noindent where $\hat{\psi}_{0,1}(\omega)$ is given by \Eqn{}~\ref{morlet2} with $r=0$ and $s=1$.  If the original input signal $x(r)$ was real, simply take the real part of \Eqn{}~\ref{InverseWavelet}.  

The constant $C_\delta$ is calculated only once for a given mother wavelet (i.~e., in the case of the Morlet, for a given value of $\omega_0$).  The discretized form of \citet{TC98} includes a different method of calculating $C_\delta$, which we will not restate here.  Using their method, we find empirically for $\omega_0 \geq 6$, 
\begin{equation}
C_\delta^{\mathrm{TC98}} = 0.776 \left( \frac{\omega_0}{6} \right)^{-1.024} , 
\end{equation}
\noindent with the uncertainty in the last decimal place.  This result allows the inverse wavelet transform to be calculated with \citeauthor{TC98}'s software, even when $\omega_0 \neq 6$.  

\subsection{Wavelet Energy and Phase}
Because our wavelet transform is complex-valued, it can be expressed in terms of two 2-D arrays: the wavelet energy spectrum (or "scalogram") 
\begin{equation}
E_W(r,s) = |T(r,s)|^2
\end{equation}
and the wavelet phase 
\begin{equation}
\label{waveletphase}
\phi_W(r,s) = \tan^{-1}\left( \frac{\mathrm{Im}\{T(r,s)\}}{\mathrm{Re}\{T(r,s)\}} \right) .
\end{equation}
\noindent Most plots of wavelet transforms in this paper will show only the energy scalogram, which gives the relative importance of different frequencies at a given location.  

The two-dimensional array of wavelet phases can be more usefully reduced to the \textit{average wavelet phase}, $\bar{\phi}_W(r)$, weighted by the modulus, over all scales $s$ at a single location $r$.  For a quasi-sinusoidal signal, $\bar{\phi}_W(r)$ gives the local proximity to peaks and troughs, even when the sinusoid's wavelength is inconstant (as it is for density waves), and can be compared to a density wave's instantaneous phase (\Eqn{}~\ref{DWPhase}).  As with the filtering method described above, selected values of $T(r,s)$ can be set to zero before performing this calculation, in order to isolate a desired component of the signal.  

Considering each complex element of $T(r,s)$ as a vector in the real-imaginary plane, $\bar{\phi}_W(r)$ can be calculated by finding the average vector, then taking its phase.  This is done by separately summing the real and imaginary parts of $T(r,s)$ over the desired scales:
\begin{equation}
\bar{\phi}_W(r) = \tan^{-1} \left( \frac{\Sigma_s \mathrm{Im}\{T(r,s)\}}{\Sigma_s \mathrm{Re}\{T(r,s)\}} \right) ,  
\end{equation}
\noindent keeping in mind that $\pi/2 \leq \bar{\phi}_W(r) \leq 3\pi/2$ iff the real part is $<0$.  

\subsection{Wavelet Ridges \label{waveletridges}}

It can be useful to plot local maxima in the 2-D representation of wavelet information.  \textit{Wavelet ridges} are those locations in the $(r,s)$ plane for which
\begin{equation}
\label{WaveletRidgeEq}
\frac{\ud [E_W(r,s)]}{\ud s} = 0 , 
\end{equation}
\noindent and are useful for pinpointing important scales (and thus frequencies, from \Eqn{}~\ref{FourierScale}) at a given location $r$ \citep{Addisonetal02}.  

\subsection{The Cone of Influence \label{coi}}

The cone of influence (COI) is the region in the wavelet transform within which edge effects may contaminate the signal.  The size of the COI varies with the scale $s$, as a consequence of the dilation of the mother wavelet.  It is up to the user to define a COI with which he or she is comfortable \citep{Addison02}. 
We follow \citet{TC98} in using the distance over which the wavelet energy contributed by the edge discontinuity decreases by a factor of $e$, which is $\sqrt{2}s$ for the Morlet wavelet.  

This form of the COI is, more generally, the interval over which the wavelet signal from any localized feature would be expected to spread.  For example, the wavelet transforms of many density waves in this paper show some signal extending to the left of $\rres$, a smearing effect which is the inevitable consequence of the spatial-frequency trade-offs discussed in Section~\ref{ChoosingOmega}.  The reader can use the COI to visualize the expected spatial resolution along any horizontal line in a transform plot.  

For a given Fourier wavelength $\lambda_F$, the COI width $\sqrt{2}s$ is proportional to the denominator in \Eqn{}~\ref{FourierScale}.  Thus, the COI is broader for higher values of $\omega_0$ (see \Fig{}~\ref{PanWakes}).  

In all wavelet plots in this paper, the COI is denoted by a region filled with cross-hatching.  It is sometimes absent, in cases where the radial scan has been cropped so that true edges are not shown.  

\subsection{Significance Levels}

When uncertainties in the input signal are characterized by gaussian white noise (with standard deviation $\sigma_{input}$), the normalized wavelet energy $E_W(r,s)/\sigma_{input}^2$ is distributed as $\chi^2$ with two degrees of freedom \citep{TC98}.  For our radial scans, each element is individually calculated as the weighted average of a number of image pixels, and thus possesses a roughly gaussian error estimate $\sigma_{input}$ (\Pone).\fn{For a radial scan element $\bar{x}$, which is a weighted average of pixel values $x_i$ with weights $w_i$, the uncertainty can be estimated with a form of the standard error of the mean (s.e.m), given by \citep{Bevington69}  
\begin{equation}
\sigma_{input}^2 = \frac{\Sigma_i [w_i (x_i-\bar{x})^2]}{N \Sigma_i w_i} .
\end{equation}
The methodology of \Pone{} does not allow for the above calculation; instead, we estimate $\sigma_{input}$ as 2 times the unweighted standard deviation, normalized by the square root of the local width of the scan area in pixels, which we find to be a good approximation for the scans discussed in this paper.  Since the only current role for $\sigma_{input}$ is to normalize contour plots of the wavelet transform, we judge this approximation to be sufficient for our purposes.}  Therefore, in this paper, the lowest level shown in each contour plot corresponds to values of $E_W(r,s)$ equal to $\sigma_{input}^2$.  Further contour levels are logarithmic, with three levels per order of magnitude in $E_W$.

\begin{table}[!t]
\begin{footnotesize}
\begin{center}
\caption{Parameter values (see \Eqn{}~\ref{DWEq}) for synthetic density wave discussed in Section~\ref{Synthetic}.  \label{some_wave_const}}
\vspace{0.1in}
\begin{tabular} { l c c }
\hline
\hline
Parameter Name & Symbol & Value\\
\hline
Resonance label & $m$ & 10\\
Background surface density & $\sigma_0$ & 40~g/cm$^2$\\
Resonance location & $\rres$ & 130,000~km\\
Initial phase & $\phi_0$ & 0\\
Damping parameter & $\xi_D$ & 10\\
Amplitude & $A_L$ & 1\\
\hline
\end{tabular}
\end{center}
\end{footnotesize}
\end{table}

\section{Wavelet Analysis of a Synthetic Density Wave \label{Synthetic}}

We here describe a multi-step semi-automated algorithm, using wavelet methods to measure the parameters describing a density wave.  To illustrate the utility and limitations of our method, we consider a signal whose characteristics are known beforehand.  We use the density wave curve taken from \Eqn{}~\ref{DWEq}, with parameters set equal to round numbers typical of the mid-A Ring (Table~\ref{some_wave_const}).  We note that these values allow the dimensionless radial parameter to be calculated as $\xi=C (r-\rres)$, where $C=0.147$~km$^{-1}$ (\Eqn{}~\ref{Def_Xi}).  

\begin{figure}[!t]
\begin{center}
\includegraphics[width=16cm,keepaspectratio=true]{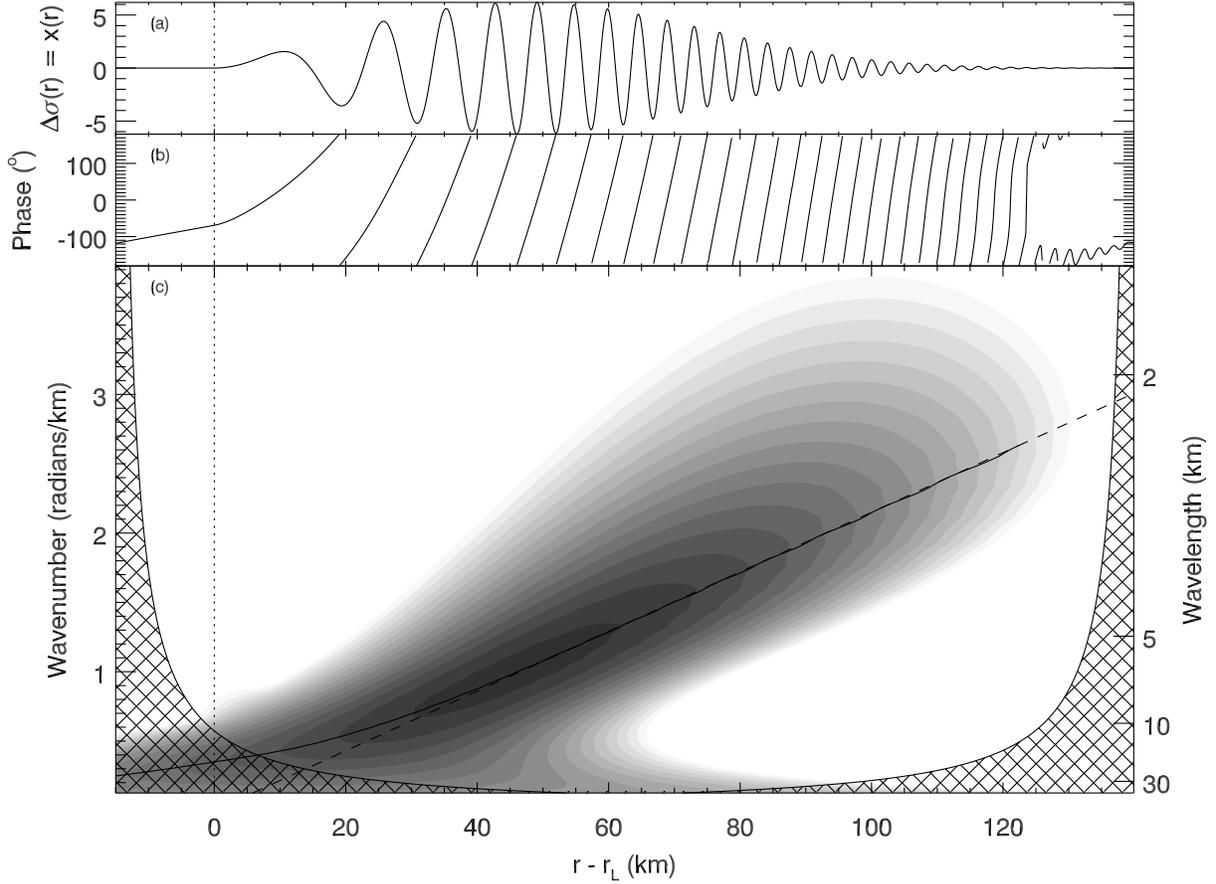}
\caption{a) Synthetic density-wave radial profile generated by \Eqn{}~\ref{DWEq} with the values of Table~\ref{some_wave_const}.  b) Wavelet phase.  c) Wavelet energy.  As with all wavelet plots in this paper, contours are logarithmic, with three contours to an order of magnitude, and the lowest contour level is at the square of the median error estimate for the scan (here we artifically set $\sigma_{input}=0.1$).  The region filled with cross-hatching is the cone of influence.  The dashed line is the foreknown wavenumber $k_{DW}(r)$ (\Eqn{}~\ref{DWWavenum}), while the solid line shows the calculated location of a wavelet ridge (\Eqn{}~\ref{WaveletRidgeEq}); here, the two diverge visibly only for $r-\rres < \textrm{40~km}$, corresponding to $\xi < 6$.  \label{some_wave1}}
\end{center}
\end{figure}

\begin{figure}[!p]
\begin{center}
\includegraphics[width=16cm,keepaspectratio=true]{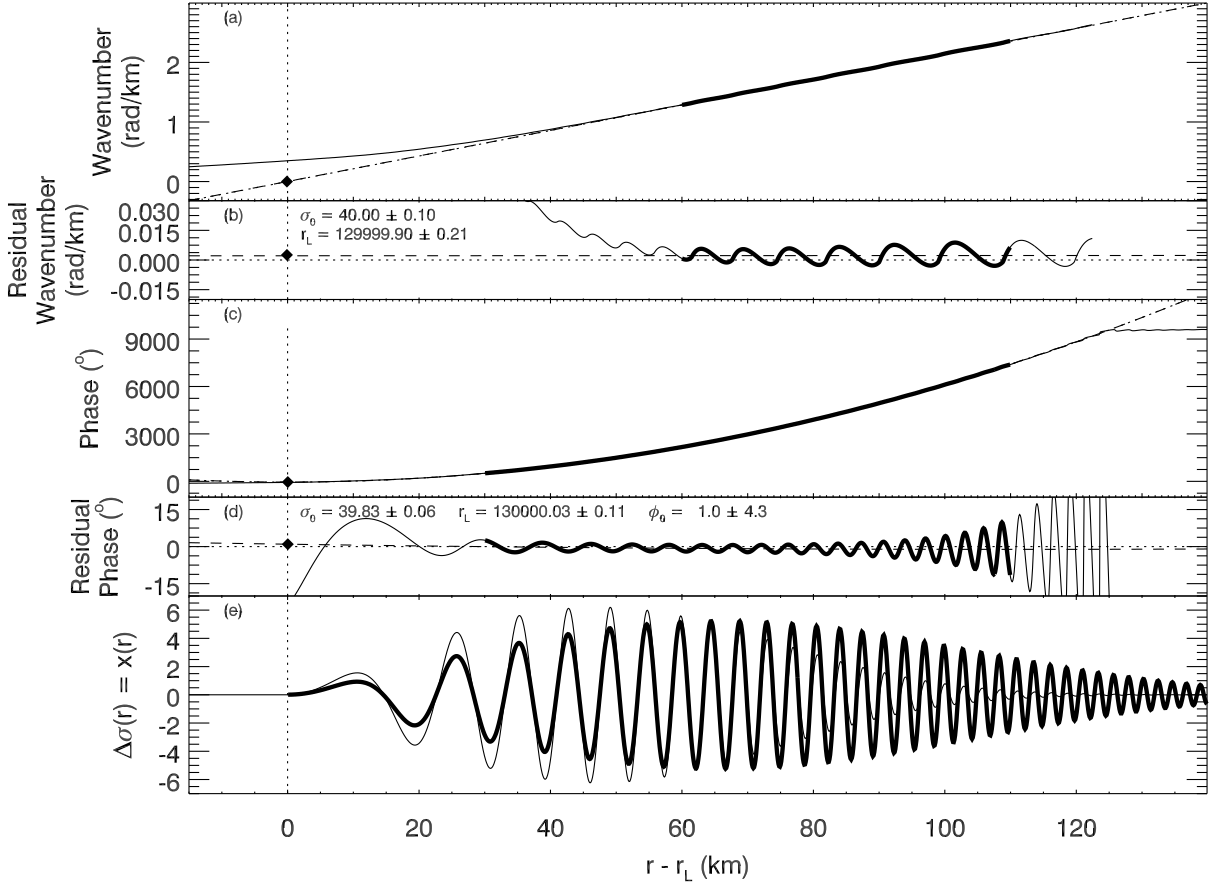}
\caption{a) Calculated wavelet ridge (Section~\ref{waveletridges}), as in \Fig{}~\ref{some_wave1}c (solid line); the expected wavenumber $k_{DW}(r)$ (dotted line).  The region of the wavelet ridge plotted in bold was used in a linear fit (dashed line); the $y=0$ point is plotted as a solid diamond.  b) All three curves from \Fig{}~\ref{some_wave3}a, shown as residuals with the expected wavenumber $k_{DW}(r)$.  c) Calculated wavelet phase $\bar{\phi}_W(r)$, as in \Fig{}~\ref{some_wave1}b, ``unwrapped'' to show how phase accumulates quadratically (solid line); the expected phase $\phi_{DW}(r)$ (dotted line).  The region of $\bar{\phi}_W(r)$ shown in bold was used in a quadratic fit (dashed line); the zero-derivative point is plotted as a solid diamond.  d) All three curves from \Fig{}~\ref{some_wave3}c, shown as residuals with the expected phase $\phi_{DW}(r)$.  e) The input synthetic density wave from \Fig{}~\ref{some_wave1}a (solid line); the fitted density wave, after the analysis of Section~\ref{HighFreq}, but still with randomly chosen values of $\xi_D$ and $A_L$ (bold solid line).  Fitted values given in the figure can be compared with the input parameters (Table~\ref{some_wave_const}) used to generated the wave.  
\label{some_wave3}}
\end{center}
\end{figure}

\subsection{Obtaining the Density Wave's High-Frequency Shape \label{HighFreq}}

\Fig{}~\ref{some_wave1} shows the radial trace of the synthetic density wave, $\Delta \sigma (r)$, from \Eqn{}~\ref{DWEq} with the values of Table~\ref{some_wave_const}, along with the phase and energy of its wavelet transform.  Note that $\Delta \sigma (r)$ as discussed in this section is identical with the generic $x(r)$ discussed in Section~\ref{WaveletTheory}.  The phase $\bar{\phi}_W(r)$ in \Fig{}~\ref{some_wave1}b is 0$^{\circ}$ at local maxima of $\Delta \sigma (r)$, and 180$^{\circ}$ at local minima, as we would expect.  \Fig{}~\ref{some_wave1}c also shows the relationship between the foreknown wavenumber (\Eqn{}~\ref{DWWavenum}) and the wavelet ridge (\Eqn{}~\ref{WaveletRidgeEq}).  For the first few wavecycles, the ridge is drawn away by non-sinusoidal components in the signal, but farther downstream it agrees with the expected wavenumber.  The residual between the two is shown in \Fig{}~\ref{some_wave3}b.  A linear fit to the portion of the wavelet ridge that is closest to linear can be used to re-obtain the background surface density $\sigma_0$ and the resonance location $\rres$ from \Eqn{}~\ref{DWWavenum}; the fitted values are also given in \Fig{}~\ref{some_wave3}b, with error estimates taken from the linear regression.  

\Fig{}~\ref{some_wave3}c shows how the wavelet phase $\bar{\phi}_W(r)$ can be ``unwrapped'' by adding 360$^{\circ}$ at every successive ``wraparound'', to reveal a shape very close to quadratic.  The residual between the wavelet phase and the expected phase $\bar{\phi}_{DW}(r)$ (\Eqn{}~\ref{DWPhase}) is shown in \Fig{}~\ref{some_wave3}d.  As with the wavelet ridge, the regions closest to $\rres$ do not correspond well to the foreknown values.  But the phase reaches stable behavior much sooner (at much lower values of $r-\rres$) than does the ridge.  As in the previous case, the portion that is closest to quadratic can be fit to re-obtain the initial phase $\phi_0$, the background surface density $\sigma_0$ and the resonance location $\rres$; the fitted values are also given in \Fig{}~\ref{some_wave3}d, with error estimates taken from the linear regression.  

We use the quadratic fit to the phase, rather than the linear fit to the wavenumber ridge, not only because it obtains all three parameters at once, but also because more of the curve is available for fitting.  In practice, it is important to look at the residuals for each individual fit, and judge those portions of the phase curve deviating from quadratic behavior, in order to exclude them; only portions that conform to quadratic behavior should be used in the fit.  

When this phase of the analysis is completed, the high-frequency behavior of the wave has been obtained, but the shape and height of the wave's envelope is still unknown, as illustrated in \Fig{}~\ref{some_wave3}e.  

\begin{figure}[!t]
\begin{center}
\includegraphics[width=16cm,keepaspectratio=true]{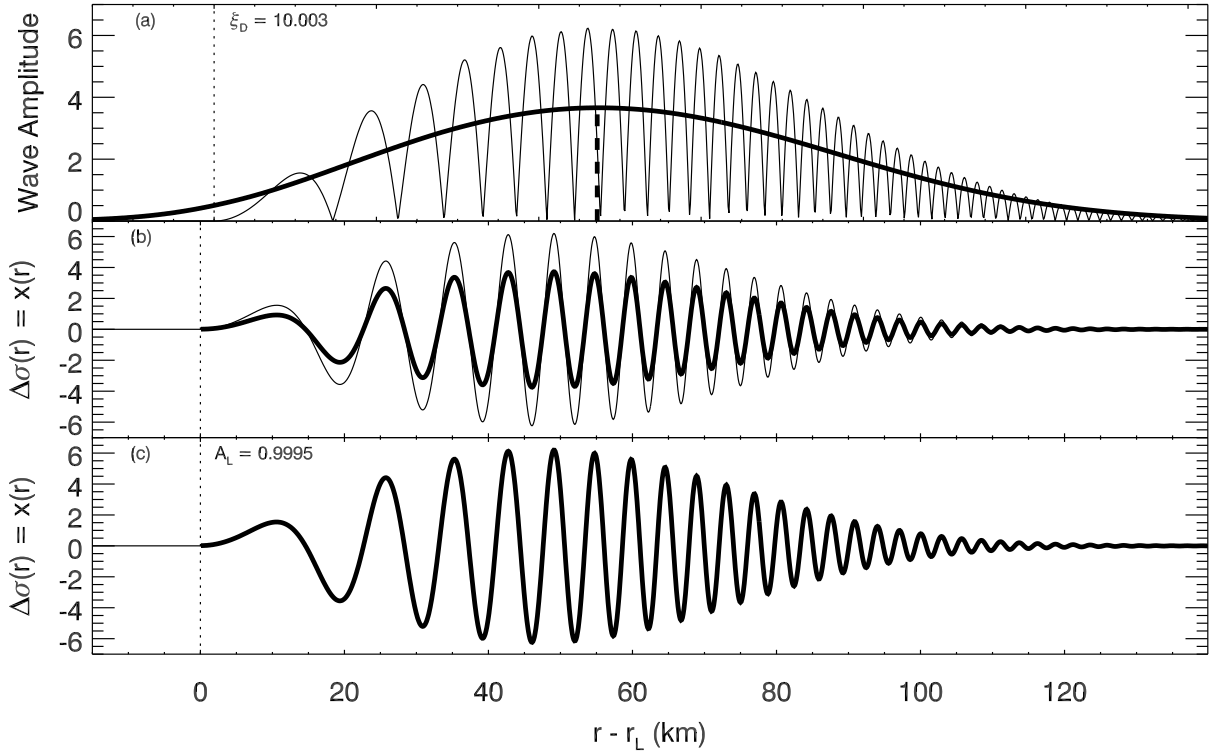}
\caption{a) The absolute value of $\Delta \sigma (r)$ (solid line); the same curve, smoothed three times with a 22-km boxcar filter, giving the low-frequency shape of the wave envelope (bold solid line); maximum point of the latter (vertical bold dashed line).  b) The input synthetic density wave from \Fig{}~\ref{some_wave1}a (solid line); the fitted density wave, after the analysis of Section~\ref{LowFreq}, but still with randomly chosen value of $A_L$ (bold solid line).  c) The input synthetic density wave from \Fig{}~\ref{some_wave1}a (solid line); the final fitted density wave, after the analysis of Section~\ref{Amp}.  Fitted values given in the figure can be compared with the input parameters (Table~\ref{some_wave_const}) used to generated the wave.  
\label{some_wave4}}
\end{center}
\end{figure}

\subsection{Obtaining the Shape of the Density Wave's Envelope \label{LowFreq}}

Now that we have determined the three parameters that define the shape of the wave's high-frequency components ($\sigma_0$, $\rres$, and $\phi_0$), we now accept that shape as given and proceed to fit the two remaining parameters ($A_L$ and $\xi_D$) that define the wave envelope (i.e.,~the amplitude and decay of the high-frequency components).  

In the absence of other effects, the shape of the amplitude modulation (i.~e., the location of the point of maximum amplitude) is governed by the damping parameter $\xi_D$.  In fact, the non-oscillatory components of \Eqn{}~\ref{DWEq} are proportional to $\xi e^{-(\xi/\xi_D)^3}$.  If a reliable representation of the amplitude can be obtained, the best way to fit for $\xi_D$ is to find the dimensionless radial parameter (\Eqn{}~\ref{Def_Xi}) at which its derivative is zero.  Denoting this point as $\xi_{max}$, we then have 
\begin{equation}
\label{Xi_max}
\xi_D = 3^{1/3}\xi_{max} .  
\end{equation}
\Fig{}~\ref{some_wave4}a illustrates the result of this calculation, in which we take the absolute value of the input waveform $\Delta \sigma (r)$ (after first ensuring that the waveform oscillates about zero) and smooth it three times with a boxcar filter.  The boxcar's width should be $\gtrsim3$ times the largest peak-to-peak wavelength in the scan.  

When this phase of the analysis is completed, only the overall amplitude of the wave remains unknown, as illustrated in \Fig{}~\ref{some_wave4}b.  

The effects of $\xi_D$ are observable only in waves that can be described throughout by the linear theory of \Eqn{}~\ref{DWEq}, and indeed the analysis in this paper is limited to such waves.  By contrast, the stronger resonances in Saturn's rings (including nearly all of the waves observed by \Voyit{}) quickly become non-linear---that is, the oscillations $\Delta \sigma$ become comparable to the background surface density $\sigma_0$, causing the assumptions underlying \Eqn{}~\ref{DWEq} to break down \citep[see][]{Shu85,BGT86}.  Non-linear density waves are characterized by sharp peaks and flat troughs, though the wavenumber $k_{DW}(r)$ is still fairly well-described by linear theory.  

Consequently, $\xi_D$ has sometimes been ignored in previous work.  Firstly, \citet{LC82} discuss and calculate the fractional distance $X_{NL}$ at which a density wave becomes non-linear, a calculation which assumes that the amplitude grows linearly with no modulation from the $\xi_D$ term.  Secondly, \citet{Rosen91a} entirely ignored the $e^{-(\xi/\xi_D)^3}$ term in \Eqn{}~\ref{DWEq} in their fits of density waves, which they justified by noting that they fit only the first few wavecycles of their waves.  It is important to recognize that both of these treatments apply only to strongly non-linear waves, and are too simplistic for the myriad of weaker waves observable with \Cassit{}'s higher resolution.  

\subsection{Obtaining the Density Wave's Amplitude \label{Amp}}

Finally, we calculate $A_L$ using a Levenburg-Marquardt least-squares fit \citep{NumericalRecipes} to compare the density wave's radial trace to the shape defined by $\sigma_0$, $\rres$, $\phi_0$, and $\xi_D$, obtained using the processes described above.  \Fig{}~\ref{some_wave4}c shows the fitted curve overlying the original synthetic wave, with the fitted parameter value also given.  

We note that it is possible to obtain $A_L$ and $\xi_D$ simultaneously with the least-squares fit, comparing the density wave's radial trace to the shape defined by only $\sigma_0$, $\rres$, and $\phi_0$.  Although this method yields acceptable results, our analysis of the synthetic wave shows that obtaining $\xi_D$ independently is superior.  Still, the two-parameter least-squares fit is occasionally necessary, such as when the wave is truncated by an image edge or by unrelated nearby radial structure.  

When the amplitude $A_L$ is given in terms of the perturbation in surface density, the mass of the perturbing moon can be obtained through appropriate coefficients from the Disturbing Function \citep[\Pone]{GT82,Rosen91a,Rosen91b}.  Unfortunately, conversion from the measured brightness ($I/F$) to optical depth, and thence to surface density, is greatly complicated by the presence of self-gravity wakes and by the ambiguities inherent in viewing the unlit side of the rings (see Section~\ref{ContrastRegime}).  As the photometric calibration is still incomplete, this paper focuses on the results obtained from the spatial information in the data (we do assume that brightness varies relatively smoothly with surface density, such that we can correctly identify maxima and minima in the latter).  We leave a more thorough analysis of wave amplitudes to a future paper.  

\subsection{Fitting the Parameters of Real Density Waves \label{FitReal}}

We apply the method described in this section to the SOI imaging data presented in this paper and in \Pone, calculating parameter values for many density waves in the rings.  Two modifications of the method are required to allow for the differences between our synthetic wave and the real data.  

\begin{figure}[!p]
\begin{center}
\includegraphics[width=11cm,keepaspectratio=true]{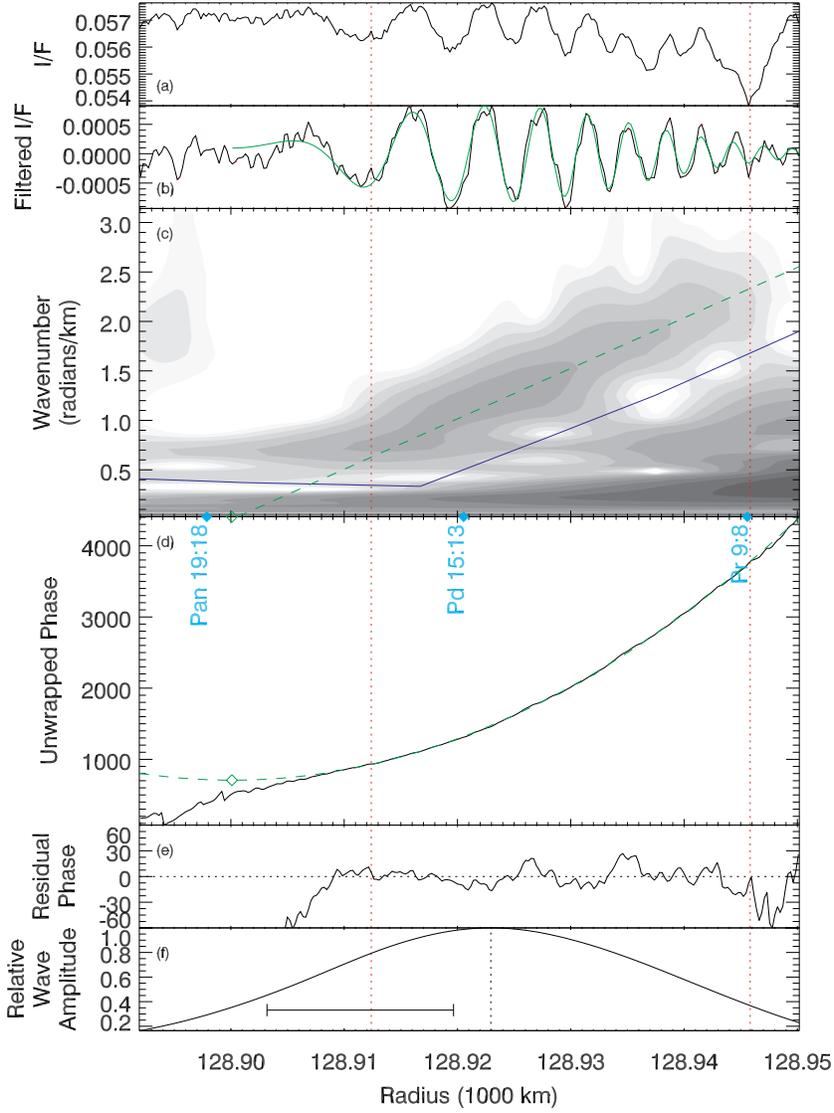}
\caption{The density wave fitting process, illustrated using the Pan~19:18 ILR density wave, observed in \Cassit{} image N1467345975 (see \Fig{s}~\ref{Prom98Image} and~\ref{SeparatingWaves2}).  The two red vertical dotted lines indicate the interval used for the quadratic fit.  From top to bottom:  a) Radial scan (\Pone{}). b) High-pass-filtered radial scan, with the final fitted wave shown in green. c) Wavelet transform of radial scan, with blue line indicating the filter boundary, and the green dashed line indicating the fitted wave's wavenumber. d) Unwrapped wavelet phase, with green dashed line indicating the quadratic fit and green open diamond the zero-derivative point. e) Residual wavelet phase, showing that the interval used for the fit is the interval in which the phase behaves quadratically. f) Wave amplitude, the local maximum of which (vertical dotted line) gives $\xi_D$; scale bar indicates the smoothing length of the boxcar filter.
\label{fit_Pan1918}}
\end{center}
\end{figure}

Firstly, we remove low-frequency components of the signal by setting selected elements of $T(r,s)$ to zero, taking the inverse wavelet transform, and then starting the analysis with this filtered wave.  This ensures that waves oscillate about zero (as does \Eqn{}~\ref{DWEq}).  In most cases, it is easy to draw the dividing line through regions of the wavelet transform with essentially zero energy (for examples, see \Fig{s}~\ref{fit_Pan1918} and~\ref{MysteryWaves2}).  More problematic cases are noted in Table~\ref{WaveFitTable}.  We do not filter out high-frequency noise, which is difficult to separate quantitatively from the desired signal. 

In thus filtering the signal, we explicitly assume that variations in the background surface density are either absent or unimportant in determining the wave structure.  In practice, we find that even waves with significant variation in background brightness can be well fit to \Eqn{}~\ref{DWEq}, thus justifying this assumption.  Waves for which such brightness variations are significant are noted in Table~\ref{WaveFitTable}.  

Secondly, we find in practice that fitted values of the initial phase $\phi_0$ are too uncertain to be useful.  They turn out to be highly correlated with $\rres$, as the distances between $\rres$ and the peaks of the wave depend linearly on $\phi_0$ \citep{Shu84}.  Furthermore, it is easy enough to calculate the value of $\phi_0$ from the location of the perturbing moon on its orbit (the uncertainty of which is exceedingly small).  Therefore, we constrain $\phi_0$ to maintain the expected value while fitting for the other parameters.  

\subsection{Error Estimates}

Meaningful uncertainties can be obtained for the parameters calculated in Section~\ref{HighFreq} --- namely $\sigma_0$, $\rres$, and $\phi_0$ --- by simply propagating the normalized errors obtained from the quadratic fit  to $\bar{\phi}_W(r)$ \citep[see][]{NumericalRecipes}.  However, this treatment assumes that the residuals will have an uncorrelated gaussian distribution about the fitted curve.  In our case this is not true.  Because the wavelet transform gives up some radial resolution in order to obtain frequency resolution (see Section~\ref{ChoosingOmega}), the effective radial resolution of the wavelet phase $\bar{\phi}_W(r)$ is lower than that of the original scan.  Thus, since we have used the same radial scale for all quantities, we have oversampled the wavelet phase so that adjacent pixels are highly correlated.  

\begin{figure}[!t]
\begin{center}
\includegraphics[width=8cm,keepaspectratio=true]{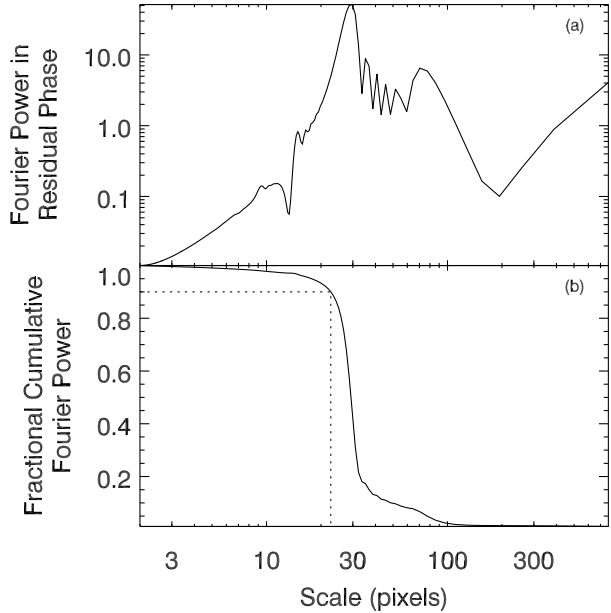}
\caption{a) The Fourier transform of the residual phase from \Fig{}~\ref{some_wave3}d.  b) The fraction of Fourier power at scales $\lambda > \lambda'$, for all $\lambda'$.  The dotted lines show that, for our synthetic wave, 90\% of the Fourier power resides at scales greater than $\lambda_{90} = 22.7$~pixels.  
\label{some_wave5}}
\end{center}
\end{figure}

We estimate the oversampling factor by examining the Fourier transform of the residual phase from \Fig{}~\ref{some_wave3}d.  Indeed, as seen in \Fig{}~\ref{some_wave5}, very little power resides at scales within an order of magnitude of the nominal pixel scale.  We define the ``effective pixel scale'' $\lambda_{90}$ as the scale above which 90\% of the Fourier power resides.  Thus, the calculated error estimates must be multiplied by a factor $\sqrt{\lambda_{90}}$.  This has been done for the synthetic error estimates quoted in \Fig{}~\ref{some_wave3}, and will be done throughout this paper.  

We estimate the uncertainty in our fitted values of $\xi_D$ as being comparable to the largest peak-to-trough wavelength in the wave, since it is these wavecycles that are smoothed over in order to estimate the shape of the wave envelope (Section~\ref{LowFreq}).  This error estimate is converted from units of km to the dimensionless units of $\xi$ by inserting it in place of the quantity $r-\rres$ in \Eqn{}~\ref{Def_Xi}.  

\subsection{Method Summary}

The steps used in this paper to fit density waves in Saturn's rings are described above in detail, illustrated in \Fig{}~\ref{fit_Pan1918}, and briefly enumerated here:  
\begin{enumerate}
\item The wavelet transform $T(r,s)$ is taken of the input signal $\Delta \sigma (r)$.
\item A line is drawn through the 2-D wavelet transform, passing through local minima, separating the wave signature from lower-frequency background variations.  
\item Elements of the wavelet transform with wavenumber below the line are set to zero, yielding the filtered wavelet transform $T'(r,s)$.  
\item The inverse wavelet transform of $T'(r,s)$ is taken, yielding a high-pass-filtered signal $\Delta \sigma'(r)$.  
\item The average wavelet phase of $T'(r,s)$ is taken, yielding $\bar{\phi}'_W(r)$.  
\item A radial interval $[r_1, r_2]$ is defined, and the phase $\bar{\phi}'_W(r)$ on $r_1 < r < r_2$ is fit to a quadratic function.  The residual between $\bar{\phi}'_W(r)$ and the fit is inspected to verify that $[r_1, r_2]$ well describes the interval that behaves quadratically.  
\item If the amplitude of $\Delta \sigma'(r)$ has a well-behaved shape (verified by visual inspection), then its maximum used to determine $\xi_D$.  
\item Taking the parameters fitted heretofore as given, a Levenburg-Marquardt least-squares fit of $\Delta \sigma'(r)$ is used to determine the remaining parameters (either $A_L$ only, or both $A_L$ and $\xi_D$).  
\end{enumerate}
Note that steps 2 and 6 explicitly require human input, while all steps are visually monitored to ensure that calculated curves are well-behaved.  

\section{Observations \label{Observations}}

\begin{figure}[!t]
\begin{center}
\includegraphics[width=16cm,keepaspectratio=true]{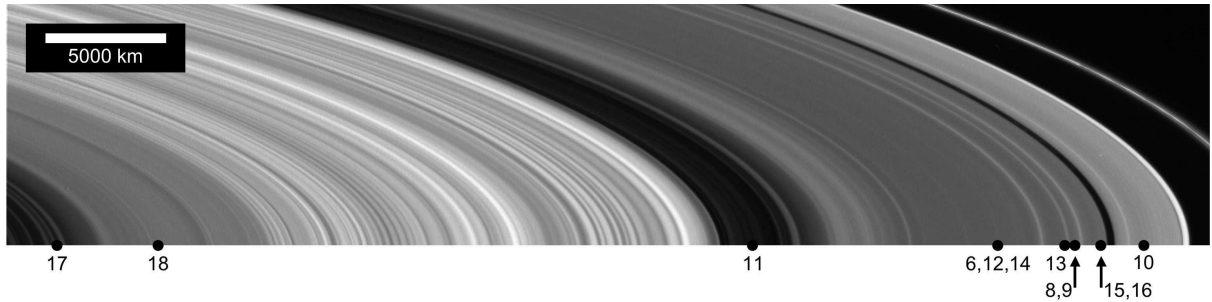}
\caption{Location within Saturn's main ring system of figures in this paper.  \label{RingsOrientation}}
\end{center}
\end{figure}

In this section we describe several highlights that emerge from wavelet analysis of the \Cassit{}~ISS radial scans of the rings (\Pone{}).  The images from which the scans used in this paper are derived are listed in Table~\ref{ObserveInfo}.  A general overview of the rings can be found in \Fig{}~\ref{RingsOrientation}, showing the locations and context of the figures that follow.  

\begin{table}[!t]
\begin{footnotesize}
\begin{center}
\caption{Observing information for images used in this paper.  \label{ObserveInfo}}
\vspace{0.1in}
\begin{tabular} { c c c c c c }
\hline
\hline
\vspace{-0.035in}
 & & Incidence & Emission & Phase & Resolution \\
Image & Date/Time (UT) & Angle$^a$ & Angle$^a$ & Angle & (m/pixel) \\
\hline
N1467344391 & 2004-183T03:15 & 114.5$^\circ$ & 46.9$^\circ$ & 82.3$^\circ$ & 180 \\ 
N1467344627 & 2004-183T03:19 & 114.5$^\circ$ & 47.0$^\circ$ & 82.1$^\circ$ & 170 \\ 
N1467345208 & 2004-183T03:29 & 114.5$^\circ$ & 62.8$^\circ$ & 59.3$^\circ$ & 270 \\ 
N1467345326 & 2004-183T03:31 & 114.5$^\circ$ & 62.9$^\circ$ & 59.3$^\circ$ & 270 \\ 
N1467345621 & 2004-183T03:36 & 114.5$^\circ$ & 62.9$^\circ$ & 59.3$^\circ$ & 250 \\ 
N1467345739 & 2004-183T03:38 & 114.5$^\circ$ & 62.9$^\circ$ & 59.2$^\circ$ & 250 \\ 
N1467345798 & 2004-183T03:39 & 114.5$^\circ$ & 62.8$^\circ$ & 59.2$^\circ$ & 240 \\ 
N1467345857 & 2004-183T03:40 & 114.5$^\circ$ & 62.8$^\circ$ & 59.2$^\circ$ & 240 \\ 
N1467345916 & 2004-183T03:41 & 114.5$^\circ$ & 62.8$^\circ$ & 59.2$^\circ$ & 240 \\ 
N1467345975 & 2004-183T03:42 & 114.5$^\circ$ & 62.9$^\circ$ & 59.2$^\circ$ & 230 \\ 
N1467346034 & 2004-183T03:42 & 114.5$^\circ$ & 62.9$^\circ$ & 59.2$^\circ$ & 230 \\ 
N1467346093 & 2004-183T03:43 & 114.5$^\circ$ & 62.9$^\circ$ & 59.3$^\circ$ & 230 \\ 
N1467346152 & 2004-183T03:44 & 114.5$^\circ$ & 62.9$^\circ$ & 59.3$^\circ$ & 220 \\ 
N1467346211 & 2004-183T03:45 & 114.5$^\circ$ & 62.9$^\circ$ & 59.3$^\circ$ & 220 \\ 
N1467346329 & 2004-183T03:47 & 114.5$^\circ$ & 62.9$^\circ$ & 59.3$^\circ$ & 210 \\
N1467351049 & 2004-183T05:06 & 114.5$^\circ$ & 94.0$^\circ$ & 132.0$^\circ$ & 770 \\
N1467351539 & 2004-183T05:14 & 114.5$^\circ$ & 94.6$^\circ$ & 135.4$^\circ$ & 840 \\
\hline
\end{tabular}
\end{center}
\begin{flushleft}
\vspace{-0.1in}
$^a$ Measured from the direction of Saturn's north pole (ring-plane normal).  Note that N1467351049 and N1467351539 view the lit face of the rings, the others the unlit face.  
\end{flushleft}
\end{footnotesize}
\end{table}

\subsection{Pan density waves \label{PanWaves}}

\begin{figure}[!t]
\begin{center}
\includegraphics[width=16cm,keepaspectratio=true]{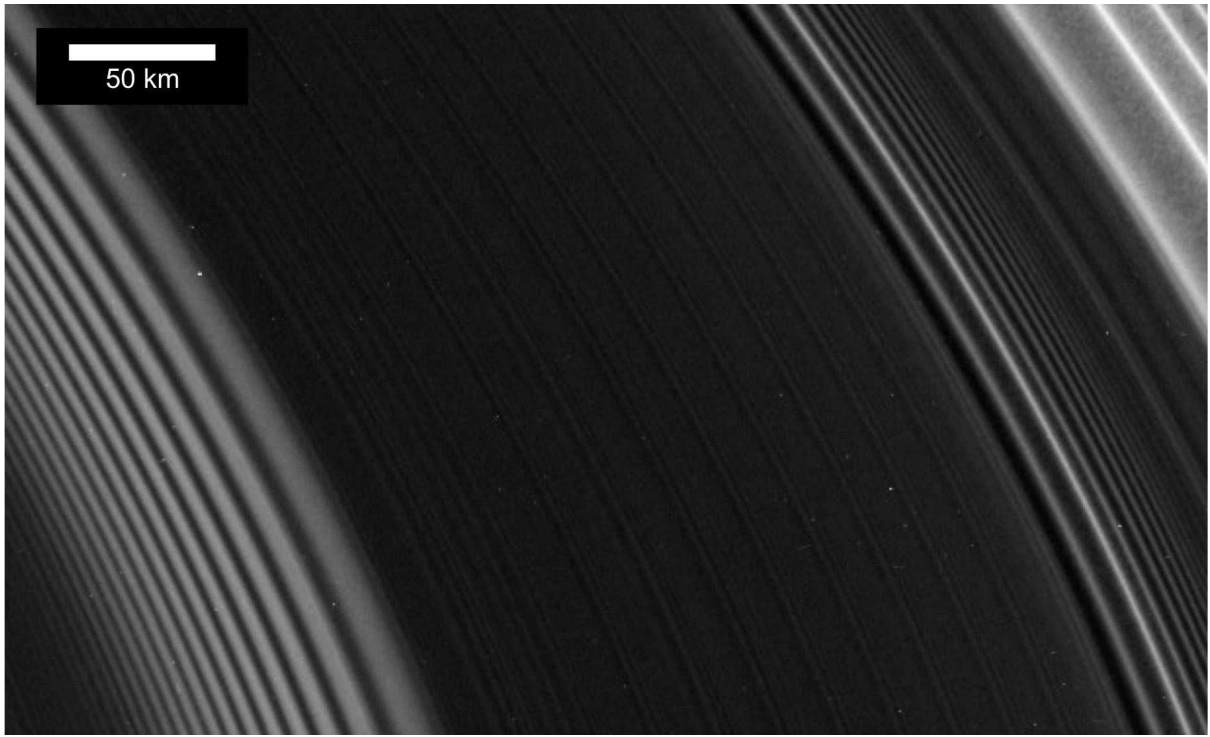}
\caption{A portion of \Cassit{} image N1467351539.  A series of small density waves, mostly due to Pan, appear like grooves on a record between the Mimas~5:3 bending wave (lower left) and the Prometheus~13:12 and Mimas~5:3 density waves (upper right).  See \Fig{}~\ref{PanSequence} for analysis.  \label{PanSequenceImage}}
\end{center}
\end{figure}

\begin{figure}[!t]
\begin{center}
\includegraphics[width=14cm,keepaspectratio=true]{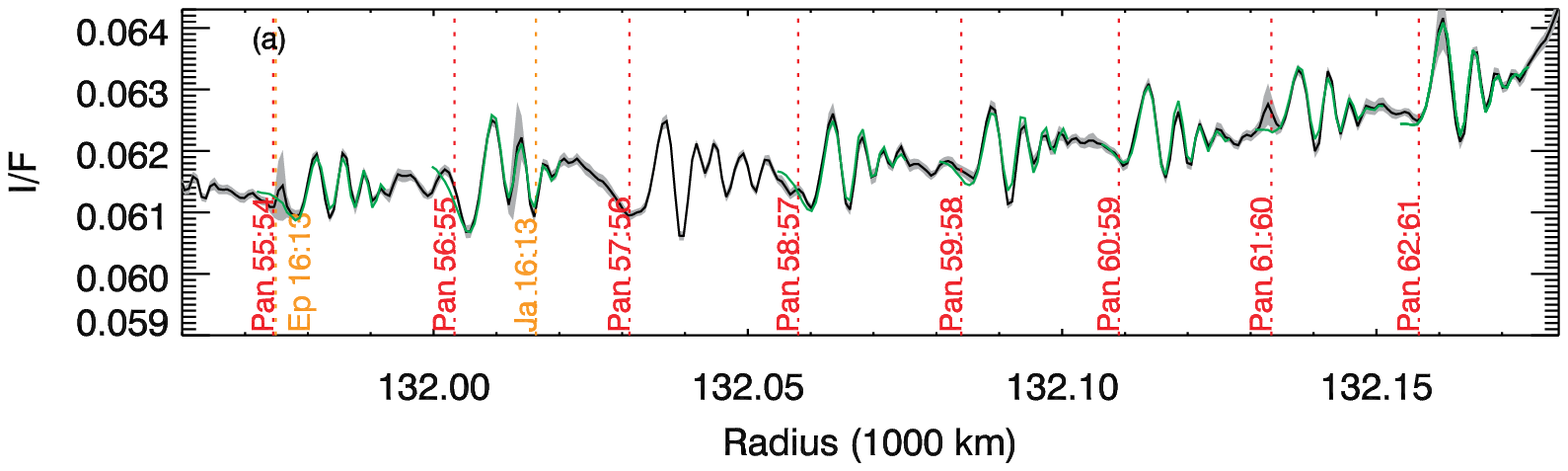}
\includegraphics[width=14cm,keepaspectratio=true]{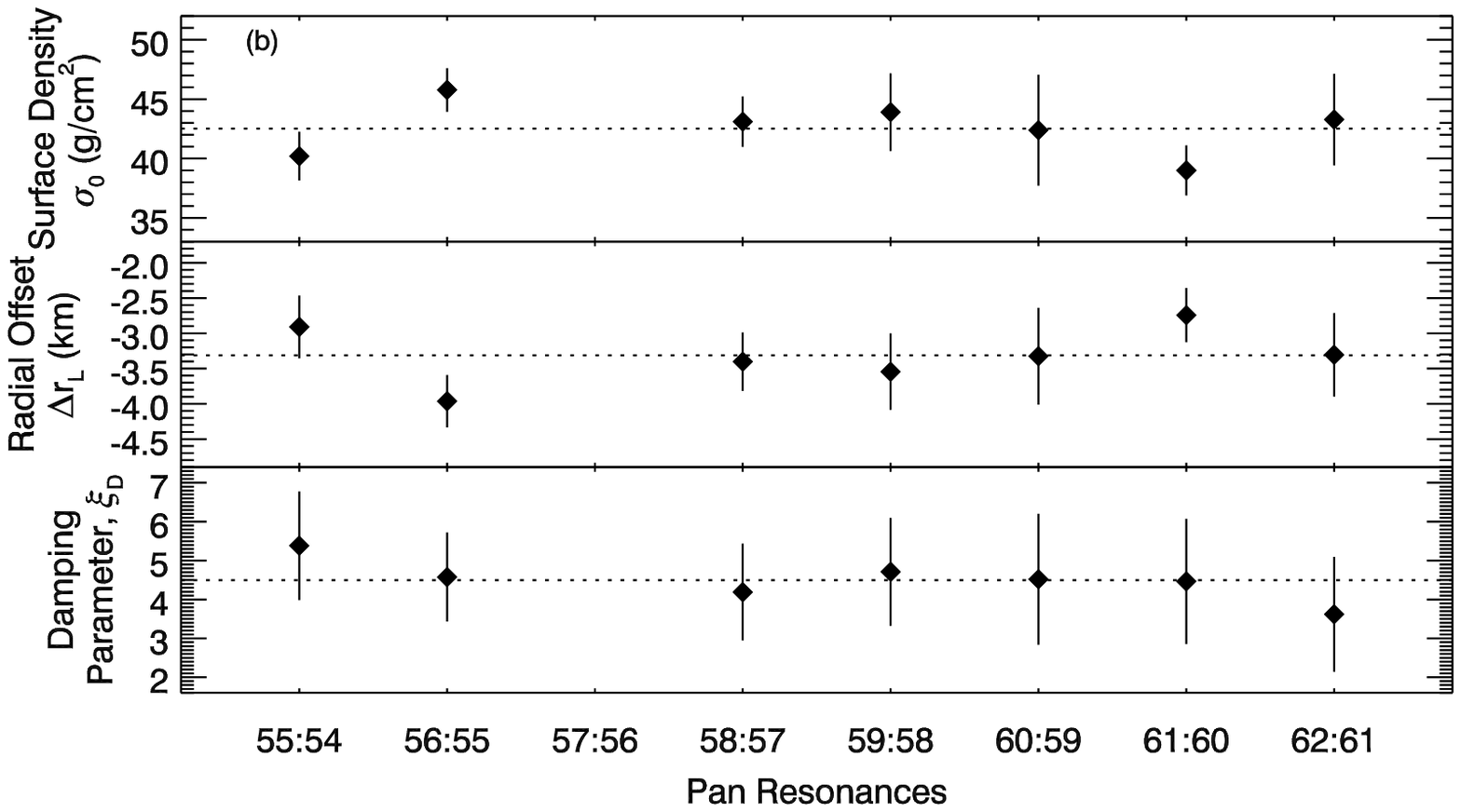}
\caption{a) Radial scan (\Pone{}) from \Cassit{} image N1467351539 (see \Fig{}~\ref{PanSequenceImage}), showing a string of density waves due to Pan ILRs 55:54 through 62:61.  Fitted waves are shown in green.  b) Three fitted parameters from these density waves, with the mean value plotted as a horizontal dotted line.  The Pan~57:56 density wave cannot be simply fit because of interference from the nearby Janus~16:13 resonance.  
\label{PanSequence}}
\end{center}
\end{figure}

The \Cassit{} ISS data set from SOI was the first to detect density waves excited by Pan, the moon embedded in the Encke Gap \citep{Porco05}; these waves are found in considerable quantity.  Because of Pan's proximity, the azimuthal parameter $m$ is limited only by the spacing between resonances becoming insufficient for a density wave to develop.  Distinct density waves with $m$ as large as 90 are observed.  The central part of \Fig{}~\ref{PanSequenceImage} (with radial trace in \Fig{}~\ref{PanSequence}a) shows a region dominated by the Pan resonances 55:54 through 62:61; fitted wave models are also shown in \Fig{}~\ref{PanSequence}a.  This region provides a cross-check on our error estimates, since adjacent waves can be assumed to have similar parameters.  \Fig{}~\ref{PanSequence}b plots the fitted parameters, showing that the error estimates well describe their mutual consistency.  

\begin{figure}[!t]
\begin{center}
\includegraphics[width=16cm,keepaspectratio=true]{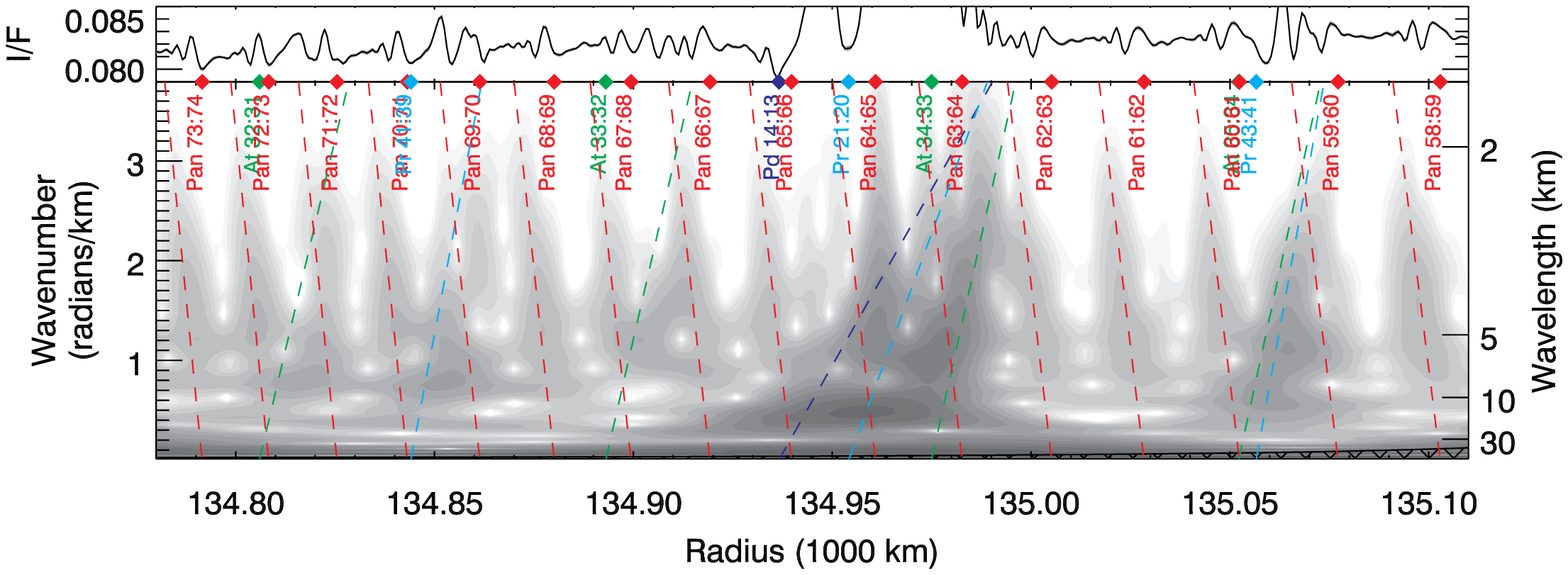}
\caption{Radial scan (\Pone{}) and wavelet transform from \Cassit{} image N1467351049, including density waves due to Pan OLRs 58:59 through 73:74, which propagate inward because they are exterior to Pan.  Density waves from other moons affect the signal at some locations.  Dashed lines indicate model density wave traces, assuming a background surface density $\sigma_0 = 15$~g/cm$^2$.  
\label{PanOLR}}
\end{center}
\end{figure}

Pan density waves are also detected exterior to the Encke Gap, constituting the first \textit{outer} Lindblad resonances (OLR) to be directly observed.  There are fewer clear examples of these, as the outer A~Ring becomes increasingly crowded with other density waves as orbital radius increases.  \Fig{}~\ref{PanOLR} shows a region of the A~Ring featuring Pan OLRs, which are unique among observed density waves in the rings in that they propagate inward.

\subsection{Atlas density waves}

The SOI data set also includes waves excited by Atlas \citep{Porco05}.  These are fewer in number than Pan density waves, as Atlas' orbit is more distant, just beyond the edge of the A~Ring.  Given its small size, Atlas' mass is much more poorly constrained by orbital integrations (in fact, \citet{Spitale06} are the first to constrain it at all), increasing the importance of measuring its mass from the amplitude of its density waves.  

\begin{figure}[!t]
\begin{center}
\includegraphics[width=16cm,keepaspectratio=true]{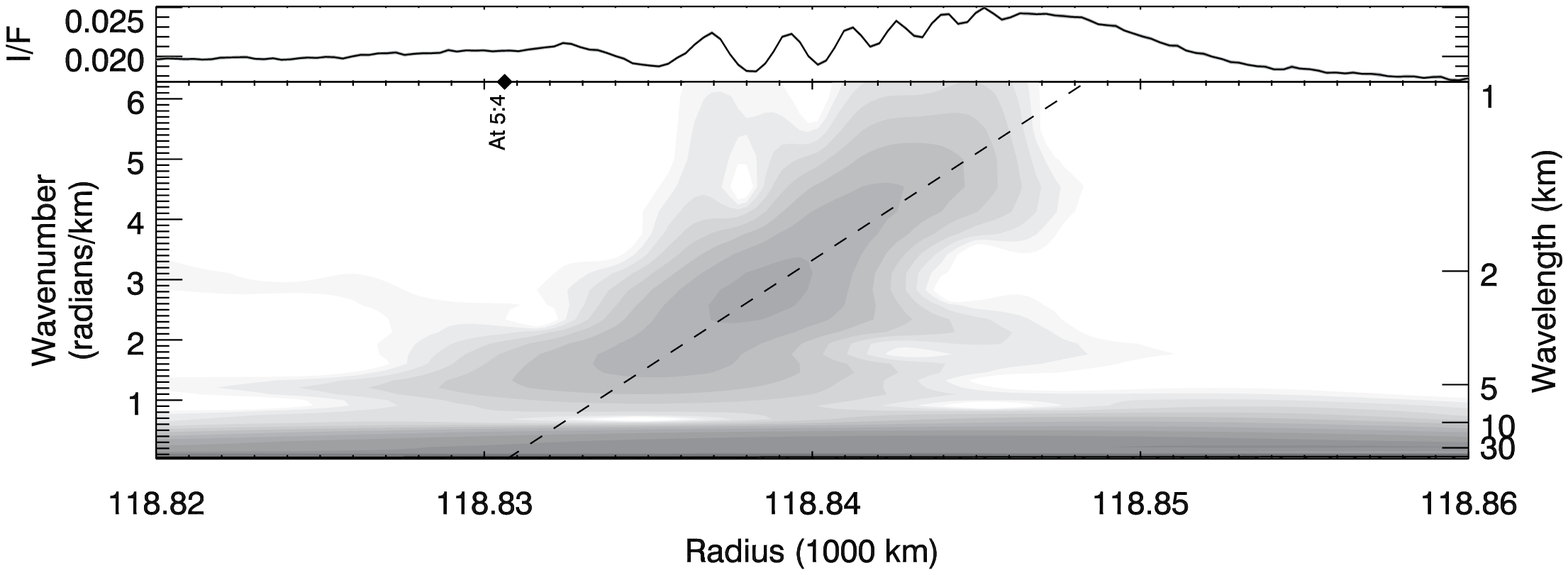}
\caption{Radial scan (\Pone{}) and wavelet transform from \Cassit{} image N1467345208, including the Atlas 5:4 density wave located in the Cassini Division.  Dashed line indicates model density wave trace, using the fitted surface density $\sigma = 1.53$~g/cm$^2$.  \label{Atlas54}}
\end{center}
\end{figure}

\Fig{}~\ref{Atlas54} shows the Atlas~5:4 density wave, which is located in a Cassini Division ringlet, and which gives the first direct measurement of surface density in that region \citep{Porco05}.  Note that this wave is perched on top of a small increase in brightness, and that the wavetrain is truncated at the point where the brightness reaches its maximum.  

\subsection{Second- and higher-order density waves \label{SecondOrder}}

Several second-order density waves excited by the larger ring moons Prometheus and Pandora are visible in our data.  Unlike the first-order waves associated with these moons, most of which were noted by \Voyit{} \citep{Rosen91a,Spilker04}, the second-order waves are weak enough that their dispersion remains strictly linear (that is, the induced variations in surface density $\Delta \sigma$ are much smaller in magnitude than the background surface density $\sigma_0$).  Thus, these waves are much better approximated by \Eqn{}~\ref{DWEq}, making them much better suited for measurements of ring parameters.  

The mass of the perturbing moon can also be inferred from these weak waves, complementing the mass estimates that have been made using orbital integrations \citep{JF04,Renner05,Spitale06}, though this must await detailed photometric modeling that can absolutely relate observed brightness to surface density. 

Second-order and third-order waves excited by the co-orbital moons Janus and Epimetheus are also present in our data.  Because of the interplay of perturbations from multiple moons, with variable orbits which periodically change the resonance locations \citep{DM81a,DM81b,Yoder83}, these structures are generally too complex to be fit using the methods described in this paper.  They are discussed more fully in \Pthree.  

\begin{figure}[!t]
\begin{center}
\includegraphics[width=16cm,keepaspectratio=true]{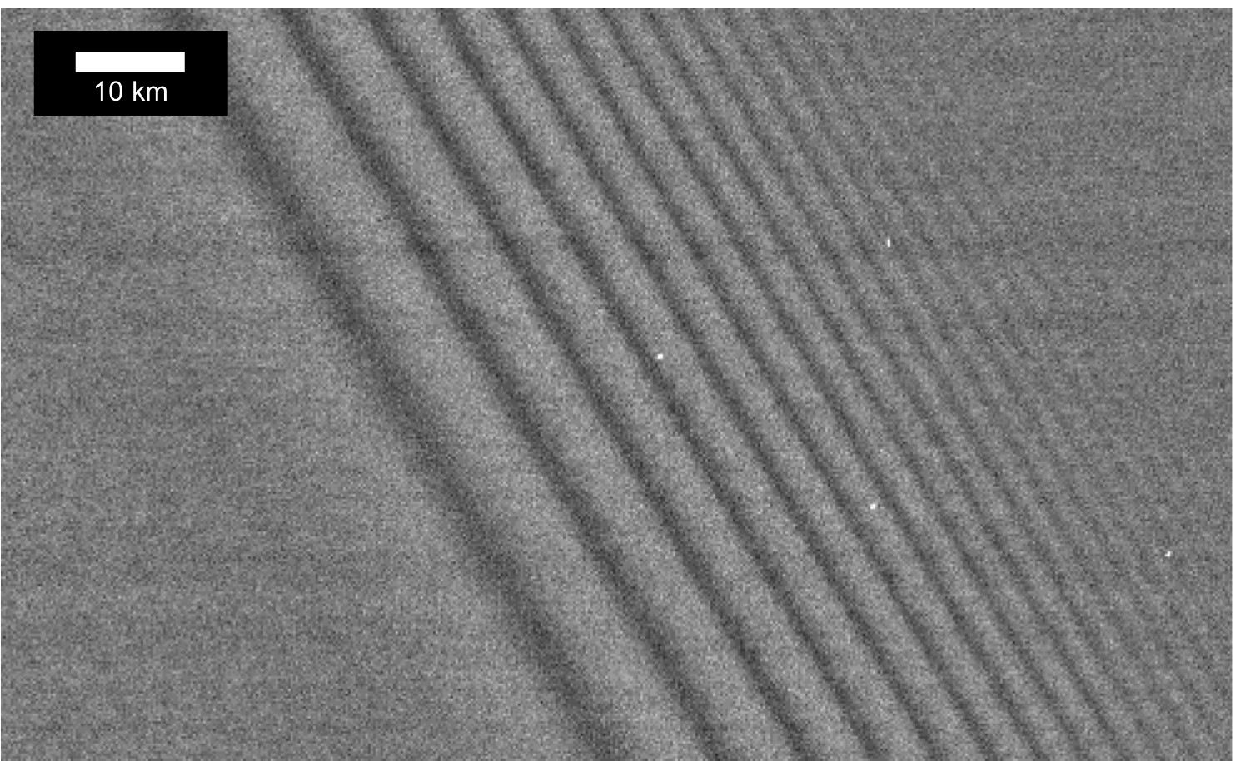}
\caption{A portion of \Cassit{} image N1467345975, showing the nearly-archetypal Prometheus~9:8 density wave.  See \Fig{}~\ref{SeparatingWaves2} for analysis.  \label{Prom98Image}}
\end{center}
\end{figure}

\begin{figure}[!p]
\begin{center}
\includegraphics[width=16cm,keepaspectratio=true]{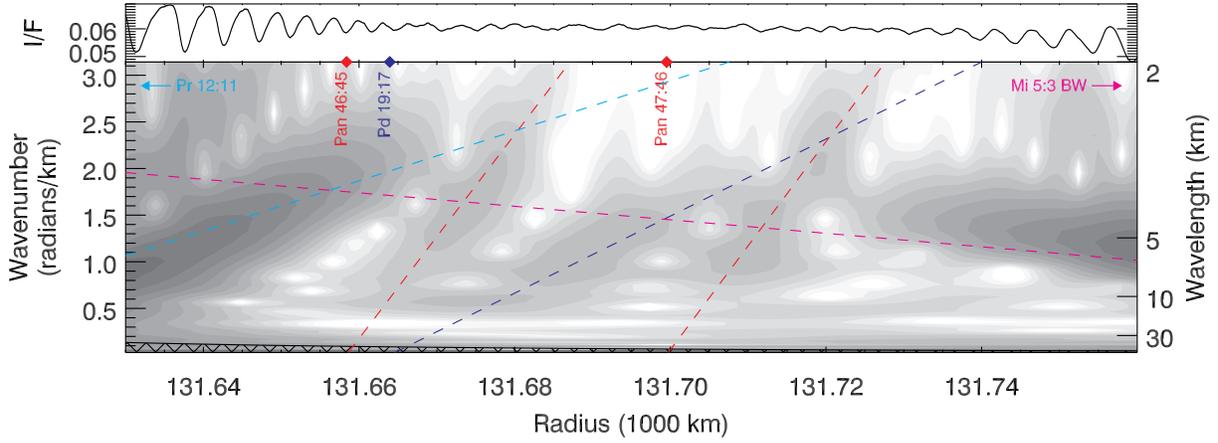}
\caption{Radial scan (\Pone{}) and wavelet transform from \Cassit{} image N1467346211.  A seemingly chaotic region between the strong Prometheus~12:11 density wave (cyan dashed line) and the very strong Mimas~5:3 bending wave (purple dashed line) is in fact populated by three distinct weak density waves: Pan~46:45 and 47:46, and Pandora~19:17.   Dashed lines indicate model density wave traces, assuming a background surface density $\sigma_0 = 37$~g/cm$^2$.  \label{SeparatingWaves1}}
\end{center}
\end{figure}

\begin{figure}[!p]
\begin{center}
\includegraphics[width=16cm,keepaspectratio=true]{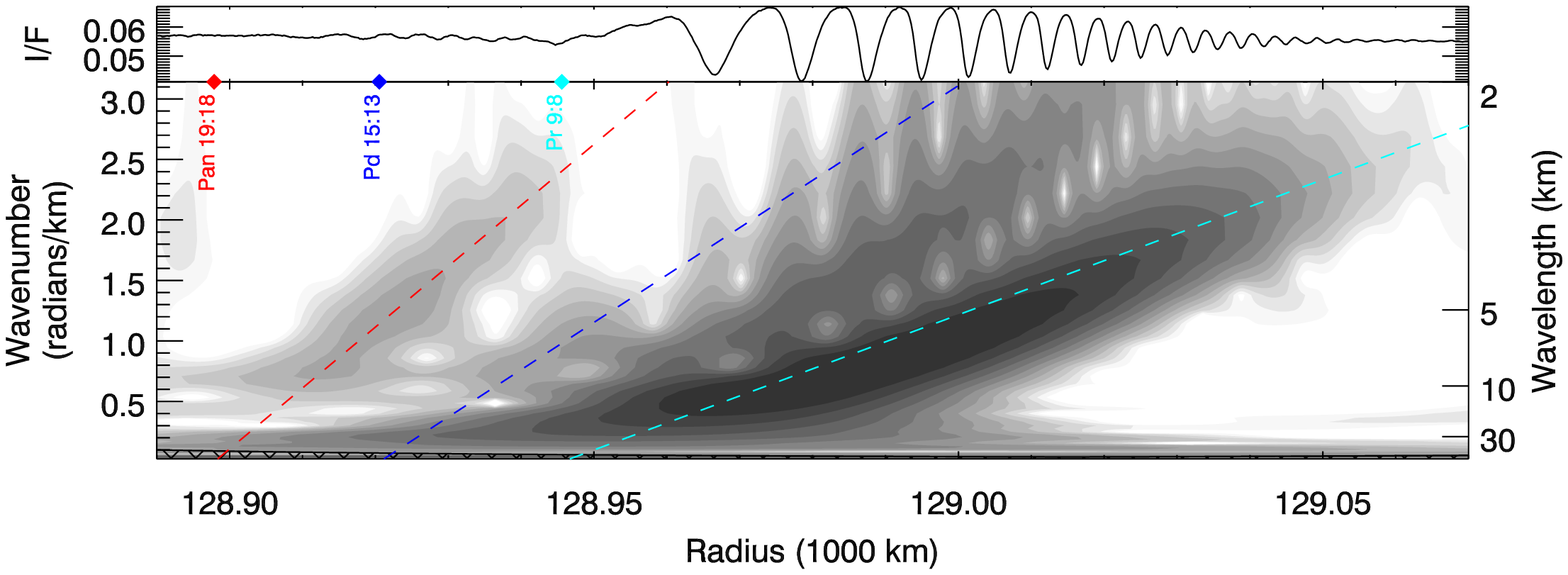}
\caption{Radial scan (\Pone{}) and wavelet transform from \Cassit{} image N1467345975 (see \Fig{}~\ref{Prom98Image}), including the weak density waves Pan~19:18 and Pandora~15:13 perched on the inward slope of the stronger Prometheus~9:8.  Dashed lines indicate model density wave traces, assuming a background surface density $\sigma_0 = 35$~g/cm$^2$.  \label{SeparatingWaves2}}
\end{center}
\end{figure}

\subsection{First-order density waves \label{FirstOrder}}

Our data show that all first-order density waves raised by the main ring-moons (i.~e., excepting the small satellites Pan and Atlas) show quantitative signs of non-linearity (see Section~\ref{LowFreq}).  The closest to linear are a handful of Prometheus and Pandora waves in the inner A Ring; in particular, Prometheus~9:8 (\Fig{s}~\ref{Prom98Image} and~\ref{SeparatingWaves2}) appears to the naked eye as a beautiful textbook example of a linear density wave (compare with the synthetic wave in \Fig{}~\ref{some_wave1}a).  However, upon closer examination, even this wave has relatively flattened brightness maxima and relatively ``peaky'' brightness minima.  In addition to being discernible in the signal, this ``peakiness'' is visible in the wavelet transform as a series of vertical streaks, as the signal becomes less like a modulated sinusoid and more like a series of $\delta$-functions (the Fourier transform of which has equal power at all frequencies).  

Another indication of non-linearity is that the wavenumber deviates from a linear function of orbital radius, instead trending towards longer wavelengths very close to resonance, and shorter wavelengths downstream.  This gives the wave a ``concave up'' appearance in wavelet space, closely resembling the behavior of the wavenumber in the non-linear wave models of \citet{BGT86} and \citet{LB86}, who show that it is due to the wave itself modulating the background surface density.  However, the middle part of the wave, especially for those waves that only barely enter the non-linear regime, may still reflect the background surface density.  A few such waves, including Prometheus~9:8, are fit using the methods of this paper.  We find that the residuals of the middle part of these waves, with respect to the quadratic fit, are low, and thus we list them as part of Table~\ref{WaveFitTable}---though, among the fitted values, only $\sigma_0$ and $\xi_D$ are likely to be very meaningful.  
 
\subsection{Separating Waves in the Frequency Domain}

\Fig{}~\ref{SeparatingWaves1} shows the region between the Prometheus~12:11 density wave and the Mimas~5:3 bending wave.  Although no coherent structure is apparent in the radial scan of this region, wavelet analysis clearly reveals three distinct smaller waves: two Pan ILRs and a second-order Pandora ILR.  

In \Fig{}~\ref{SeparatingWaves2}, wavelet analysis of the radial trace of the Prometheus~9:8 density wave reveals that its inward regions are convolved with the second-order Pandora~15:13 density wave, as well as the Pan~19:18 density wave.  

Although these waves are not well enough separated to be quantitatively fitted by the method described in Section~\ref{Synthetic}, they illustrate the ability of wavelet analysis to separate out and identify waves invisible to the naked eye.  Yet another example is the identification of the very weak Janus~14:11 density wave in \Fig{}~\ref{PanWakes}

\begin{figure}[!t]
\begin{center}
\includegraphics[width=16cm,keepaspectratio=true]{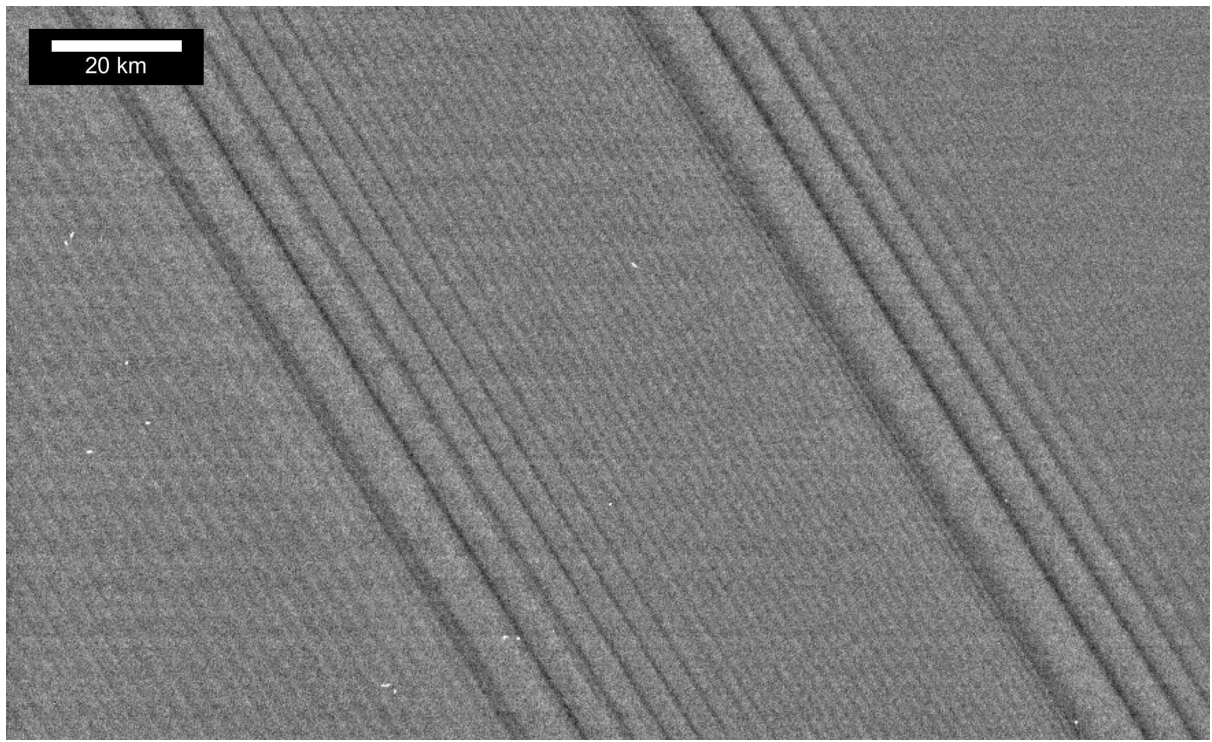}
\caption{A portion of \Cassit{} image N1467346329, showing a ``corduroy'' pattern caused by moonlet wakes excited by Pan.  Also seen are the Pandora~11:10 and Prometheus~15:14 density waves.  See \Fig{}~\ref{PanWakes} for analysis.  \label{PanWakesImage}}
\end{center}
\end{figure}

\begin{figure}[!t]
\begin{center}
\includegraphics[width=16cm,keepaspectratio=true]{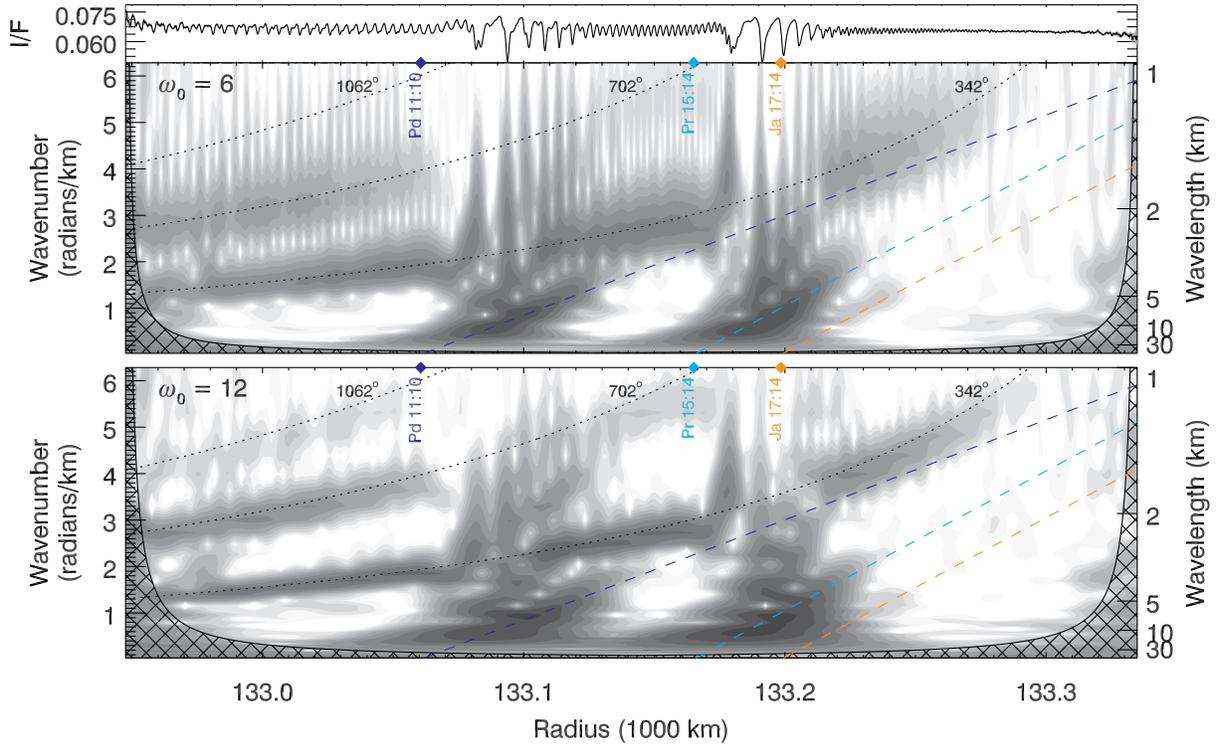}
\caption{Radial scan (\Pone{}) and two wavelet transforms from \Cassit{} image N1467346329 (see \Fig{}~\ref{PanWakesImage}), showing wakes excited by Pan along with several density waves.  The upper wavelet plot, like all others in this paper, uses a central frequency $\omega_0 = 6$; the lower plot uses $\omega_0 = 12$, resulting in increased resolution in the spectral ($y$) dimension at the expense of smearing in the radial ($x$) dimension (see Section~\ref{ChoosingOmega}).  Density waves are clearer in the upper plot, including the strong Pandora~11:10 and Prometheus~15:14 waves, but also the weak third-order Janus~17:14.  Dashed lines indicate model density wave traces, assuming a background surface density $\sigma_0 = 40$~g/cm$^2$.  Moonlet wakes excited by Pan are clearer in the lower plot.  The three dotted lines denote the frequency profiles \citep{Show86} of wakes that have traveled 342$^\circ$, 702$^\circ$, or 1062$^\circ$ in synodic longitude since their last encounters with Pan.  \label{PanWakes}}
\end{center}
\end{figure}

\subsection{Pan wakes}

The ``corduroy'' pattern pervading the image in \Fig{}~\ref{PanWakesImage} is caused by moonlet wakes excited by Pan.  Such wakes (described in detail by \citet{Show86} and \citet{Horn96}) are a fundamentally different process from waves in that they do not propagate, but arise from ring particle orbits organized coherently by encounters with nearby Pan.  

Wavelet analysis of the image is shown in \Fig{}~\ref{PanWakes}, with two different values of the central frequency $\omega_0$ (Section~\ref{ChoosingOmega}) bringing out different aspects of the structure.  Like previous authors, we note that wake wavelengths in the data tend to be a few percent longer than the model predicts, likely due to mutual gravitation among the ring particles.  Secondly, not only can we verify the existence of a second-order wake (which is simply structure that has survived after more than one synodic period after its creation by Pan), but we also see signs of a third-order wake.  Thus, wake structure is seen right down to wavelengths near the resolution limit, indicating that we still cannot place a limit on the remarkable resiliency of wake structures against damping.  

\begin{figure}[!t]
\begin{center}
\includegraphics[width=16cm,keepaspectratio=true]{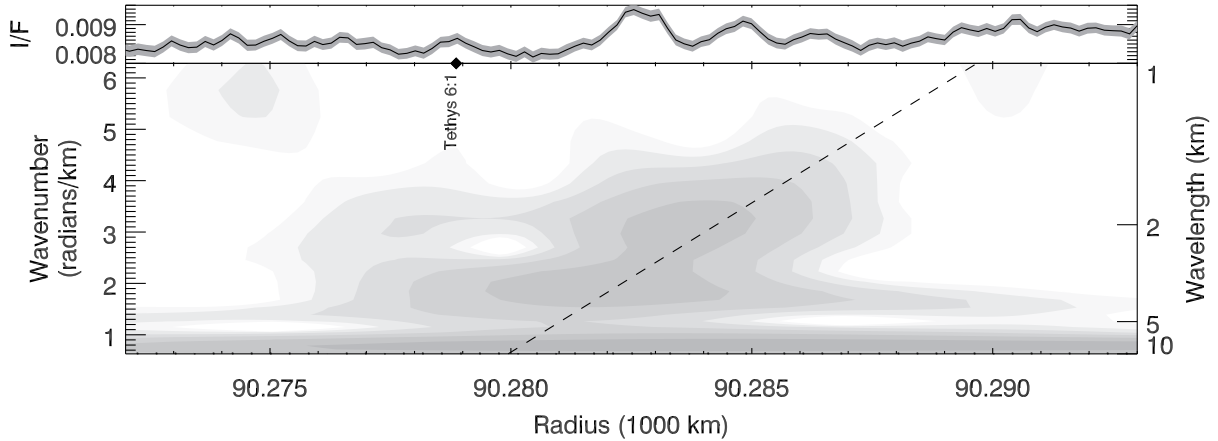}
\caption{Radial scan (\Pone{}) and wavelet transform from \Cassit{} image N1467344391, showing an unidentified wave-like feature in a non-plateau region of the C~Ring.  Although the Tethys~6:1 ILR is likely too weak to excite observable perturbations, its location (fixed by a fiducial feature elsewhere in the image) is intriguing.  Dashed line indicates model density wave trace, assuming a background surface density $\sigma_0 = 0.7$~g/cm$^2$  \label{MysteryWaves1}}
\end{center}
\end{figure}

\subsection{Unidentified Waves}

Several examples exist in the SOI data set of structures that appear to be waves but are unidentified.  

The first is shown in \Fig{}~\ref{MysteryWaves1}, taken from an image of a non-plateau region of the C~Ring.  The Tethys 6:1 ILR at 90,279~km presents itself as an intriguing explanation for this wave-like structure.  The surface density obtained for an $m=2$ wave is $\sigma_0 \sim 0.7$~g/cm$^2$, which is lower than any other direct measurement of ring surface density, but which may be consistent with the low optical depth of this region.  Unlike most images considered in this paper, the radial scale here is constrained by a nearby circular feature at 90,405.7~km (designated ``IER~24'' by \citet{French93}), so the spatial correlation between the resonance and the observed feature is not artificial.   However, the calculated strength (torque) of the Tethys 6:1 ILR is several orders of magnitude lower than many resonances observed to excite no waves (\Pone{}).  

\begin{figure}[!t]
\begin{center}
\includegraphics[width=16cm,keepaspectratio=true]{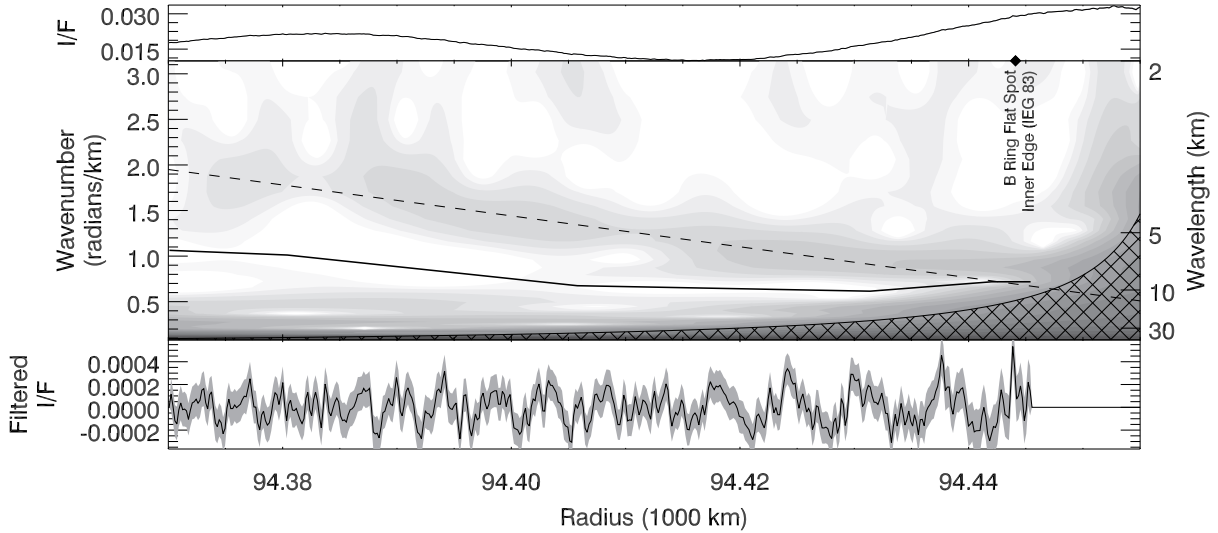}
\caption{Radial scan (\Pone{}) and wavelet transform from \Cassit{} image N1467344627, showing an unidentified wave-like feature in the B~Ring.  Dashed line indicates model density wave trace, assuming a background surface density $\sigma_0 = 60$~g/cm$^2$, a resonance location $\rres = 94,485$~km, and a azimuthal parameter $m = 4$.  Solid line indicates a high-pass filter boundary, and lower panel shows the radial scan obtained by inverting the filtered wavelet transform.  \label{MysteryWaves2}}
\end{center}
\end{figure}

\Fig{}~\ref{MysteryWaves2} shows a second unidentified feature, at the inner edge of the relatively featureless region of the inner B Ring known as the ``Flat Spot'' (designated ``IEG~83'' by \citet{French93}). Superimposed upon the $\sim 100$~km undulation is a quasi-oscillatory signal whose frequency increases linearly with \textit{decreasing} orbital radius, reminiscent of an inward-propagating density or bending wave.  Assuming a background surface density $\sigma_0 \sim 60$~g/cm$^2$, obtained from the nearby Janus~2:1 density wave, we estimate the azimuthal parameter $m \sim 4$.  

\subsection{Photometric Regime \label{ContrastRegime}}

Neither ring surface density nor optical depth is directly observed in imaging data; rather, these quantities must be obtained through photometric modeling of the directly-observed brightness \citep[e.~g.,][]{Dones93}.  This is in contrast to radio and stellar occultations, which directly measure the optical depth.  Detailed models to obtain optical depths from \Cassit{} imaging data are not yet complete, which is why this paper focuses on the spatial information in the data, leaving analysis of wave amplitudes for the future.  For this purpose, all that is necessary is to know whether maxima (minima) in brightness correspond to maxima (minima) in surface density, or vice versa.  The former is the ``normal-contrast'' regime; the latter is ``reverse-contrast.''  

\begin{figure}[!t]
\begin{center}
\includegraphics[width=8cm,keepaspectratio=true]{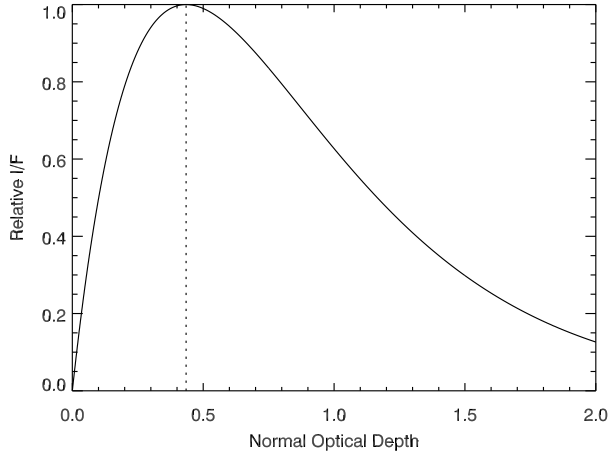}
\caption{Ring brightness (I/F) as a function of normal optical depth ($\tau$), under the assumption of single scattering, using geometrical parameters appropriate for our highest-quality images (solar incidence angle 114.5$^\circ$, emission angle 62.8$^\circ$).  Vertical dotted line indicates the location of $\tau_{max}$.  \label{soirings_taumax}}
\end{center}
\end{figure}

Images of the illuminated face of the rings always have normal contrast.  However, ambiguity exists for the unilluminated face where, for example, no signal might mean either a totally clear region or a totally opaque region.  This simple illustration indicates that the maximum brightness occurs at a finite optical depth $\tau_{max}$, with normal contrast holding for optical depths $\tau<\tau_{max}$ but reverse contrast for $\tau>\tau_{max}$.  The value of $\tau_{max}$ can be estimated using the assumption of single scattering in a classical many-particle-thick ring; the dependence on optical depth is proportional to $e^{-\tau/\mu'} - e^{-\tau/\mu}$, where $\mu'$ and $\mu$ are the cosines of the solar incidence and the emission angles, respectively \citep{Chandra60,Cuzzi84}.  
For the observing geometry of the highest-quality images considered here, $\tau_{max} \sim 0.43$ (\Fig{}~\ref{soirings_taumax}).  

Mean background optical depths for the A~Ring were measured by \Voyit{} at $\tau \gtrsim 0.5$ \citep[e.g.][]{Espo83}, leading one to expect reverse contrast in the unlit-side images of the A~Ring.  However, recent studies \citep{Colwell06,Hedman07} indicate that, rather than being a homogeneous sheet at the mean optical depth, the A~Ring is instead an ``intimate mixture'' of nearly opaque self-gravity wakes with much more diffuse material ($\tau_{gap} \sim 0.15$) in the gaps between the wakes.  The latter may well dominate the scattering of light into the camera, putting the images in the normal-contrast regime.  A second argument in favor of the normal-contrast regime is that it results in lower error estimates in our model fits.\fn{The ratio in the uncertainties of $\sigma_0$ and $\rres$ (the contrast ambiguity does not affect $\xi_D$ or $A_L$) between reverse-contrast and normal-contrast ranges from 0.8 to 6.1, with a geometric mean of 1.5.}  

However, we adopt here the classical assumption of reverse contrast for this observing geometry.  The morphology of non-linear density waves in the high-resolution images (Section~\ref{FirstOrder}), the brightness minima of which are ``peaky'' while the brightness maxima are flattened \citep[see][]{Shu85}, is most naturally interpreted with the reverse-contrast assumption.  The morphology of the ``straw'' texture observed in strong density waves \citep{Porco05} similarly favors reverse-contrast.  Finally, the surface densities derived under the reverse-contrast assumption agree better with the work of previous authors, with waves from the lit-side (but lower-resolution) images reported here, and with preliminary analysis of more recent high-resolution lit-side data (not reported here).  

\section{Results of Wavelet Analysis \label{Results}}

\begin{figure}[!p]
\begin{center}
\includegraphics[width=16cm,keepaspectratio=true]{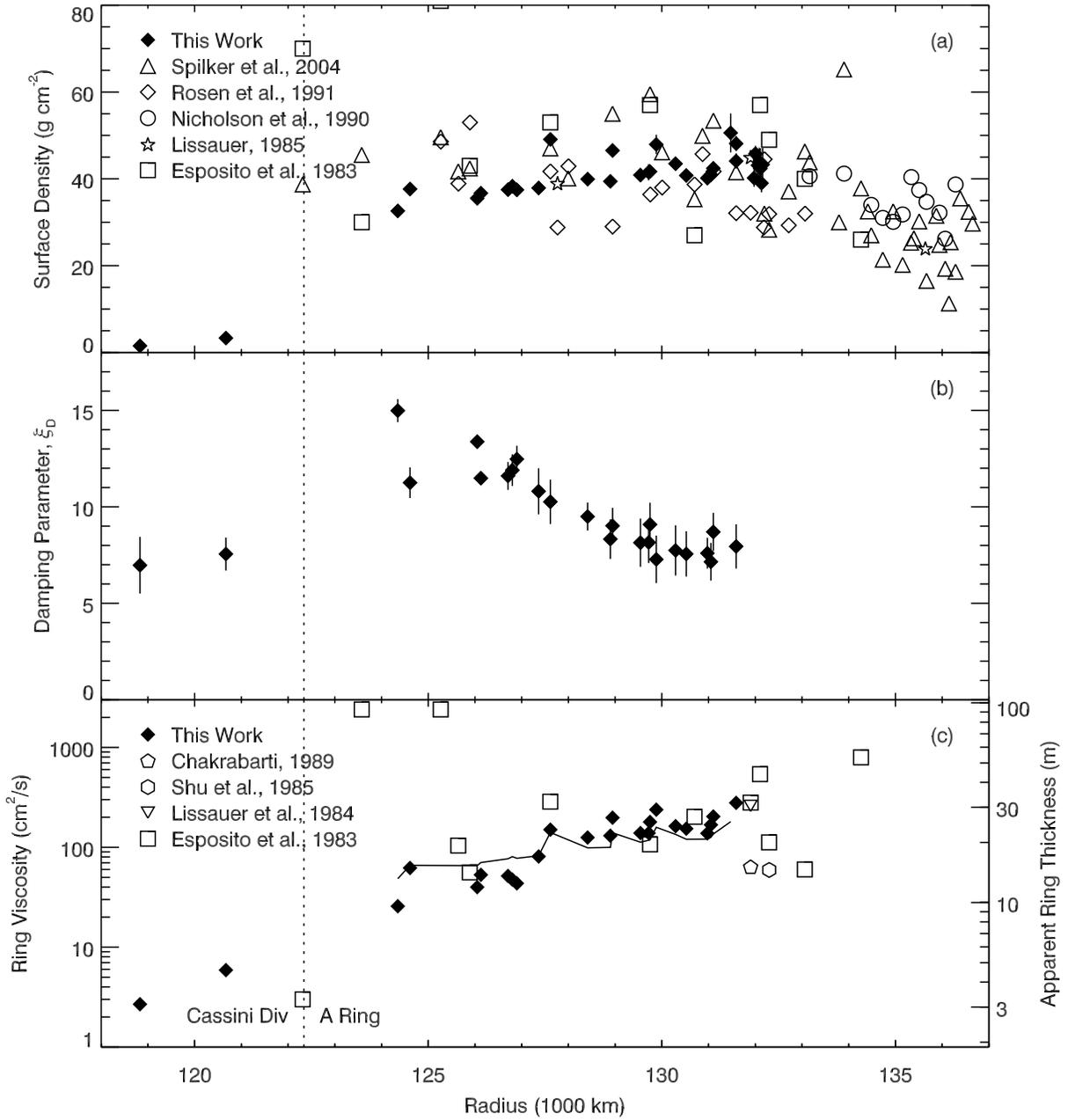}
\caption{a) Surface density $\sigma_0$ in Saturn's rings, as measured using density waves here and in previous work.  Vertical lines indicate error bars; for points with no visible error bar, the error bar is smaller than the plotted symbol.  Error bars for previous work (not shown) are generally similar in magnitude to their scatter \citep[see][]{Spilker04}.  b) Damping parameter $\xi_D$ in Saturn's rings.  Vertical lines indicate error bars.  c) Viscosity in Saturn's rings, as calculated here and in previous work.  The associated apparent ring thickness is an upper limit (see text).  Solid line indicates the theoretical prediction of \citet{Daisaka01}, as calculated from \Eqn{}~\ref{DaisakaPredict}.  \label{soirings_fitresults}}
\end{center}
\end{figure}

\begin{figure}[!t]
\begin{center}
\includegraphics[width=16cm,keepaspectratio=true]{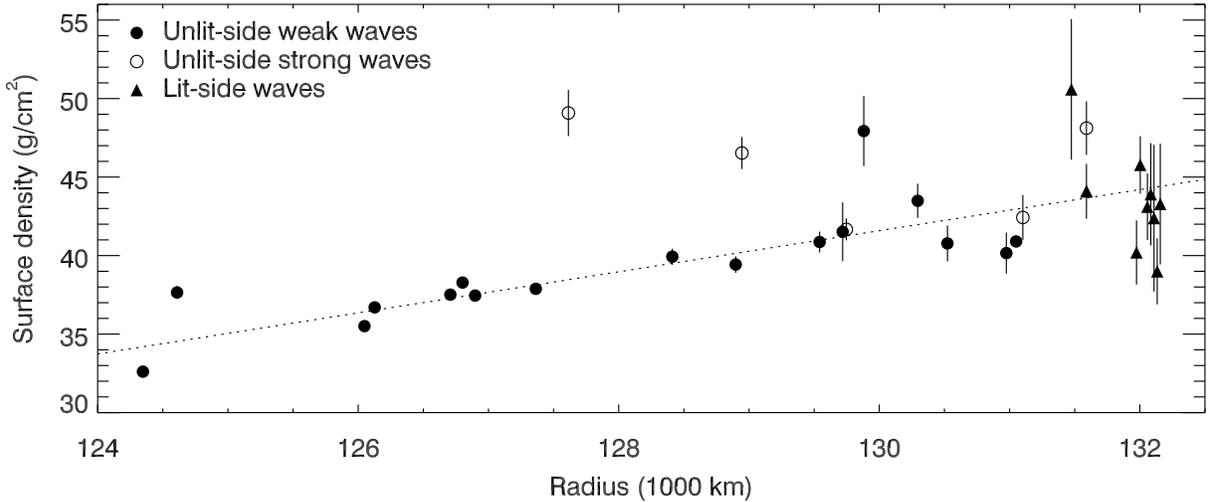}
\caption{Surface density $\sigma_0$ in the inner- and mid-A Ring for our results only (same data as in \Fig{}~\ref{soirings_fitresults}).  Symbols indicate whether a wave appears in an unlit-side or a lit-side image (the former have better resolution), and whether an unlit-side wave is weak or strong (the former are more well-behaved).  Dotted line shows linear fit $\sigma_0 = (33.7 \pm 0.1 ) + (1.3 \pm 0.5) ( r_{1000} - 124 )$, where $\sigma_0$ is in g/cm$^2$ and $r_{1000}$ is in thousands of km.  
\label{soirings_fitresults_alla}}
\end{center}
\end{figure}

The density wave parameters measured using the method described in Section~\ref{Synthetic} are given in Table~\ref{WaveFitTable} and plotted in \Fig{s}~\ref{soirings_fitresults} and~\ref{soirings_fitresults_alla}.  

\begin{table}[!p]
\begin{footnotesize}
\begin{center}
\caption{Fitted values of density wave parameters including resonance location ($\rres$); deviation of the same from nominal pointing ($\Delta\rres$); background surface density ($\sigma_0$); initial phase ($\phi_0$); damping parameter ($\xi_D$); wavelength, in pixels, at maximum amplitude ($\lambda_{max}$); and amplitude ($A_L$).  Fits assume the reverse-contrast regime (Section~\ref{ContrastRegime}).  
\label{WaveFitTable}}
\begin{tabular} { @{} l c c r @{$\pm$} l r @{$\pm$} l c r @{$\pm$} l c c @{} }
\hline
\hline
& Image & $\rres$ (km) & \multicolumn{2}{c}{$\Delta\rres$ (km)} & \multicolumn{2}{c}{$\sigma_0$ (g/cm$^2$)} & \multicolumn{1}{c}{$\phi_0$ ($^\circ$)} & \multicolumn{2}{c}{$\xi_D$} & $\lambda_{max}$ (px) & $A_L$ (I/F)\\
\hline
Atlas 5:4 & N1467345208 & 118830.62 &  -9.84 & 0.22 &  1.52 &  0.06 & 323.1 &  6.97 &  1.47$^f$ &  8.0 &  41.51\\
Pan 7:6$^a$ & N1467345326 & 120669.37 &  -7.65 & 0.11 &  3.32 &  0.04 & 201.2 &  7.55 &  0.85$^f$ &  9.4 &  15.68\\
Atlas 7:6 & N1467345621 & 124347.51 &  -4.24 & 0.30 & 32.61 &  0.21 &   0.4 & 14.99 &  0.60 & 16.6 &   9.10\\
Pan 10:9$^b$ & N1467345621 & 124609.75 &  -7.45 & 0.25 & 37.65 &  0.42 & 328.7 & 11.25 &  0.80 & 19.5 &   9.04\\
Atlas 8:7$^c$ & N1467345739 & 126048.12 &  -5.61 & 0.24 & 35.51 &  0.21 & 288.6 & \multicolumn{2}{c}{13.38$^e$} & 18.9 &   7.88\\
Pan 12:11$^c$ & N1467345739 & 126126.36 &  -6.09 & 0.18 & 36.70 &  0.22 & 293.3 & \multicolumn{2}{c}{11.48$^e$} & 17.9 &   9.80\\
Pan 13:12 & N1467345798 & 126707.39 &  -6.10 & 0.11 & 37.51 &  0.15 & 274.8 & 11.60 &  0.72 & 17.5 &   7.59\\
Prometheus 15:13$^c$ & N1467345798 & 126800.70 &  -6.72 & 0.11 & 38.28 &  0.17 & 146.1 & 11.90 &  0.82 & 16.6 &   6.57\\
Pandora 13:11 & N1467345798 & 126897.51 &  -6.26 & 0.26 & 37.45 &  0.34 & 275.6 & 12.47 &  0.69 & 17.1 &   7.46\\
Atlas 9:8$^c$ & N1467345857 & 127363.40 &  -5.30 & 0.27 & 37.88 &  0.27 & 215.8 & 10.81 &  1.20 & 23.8 &   9.79\\
Prometheus 8:7 & N1467345857 & 127613.59 & -10.15 & 1.14$^d$ & 49.08 &  1.47 & 178.8 & 10.26 &  1.15 & 30.6 & 173.90$^d$\\
Atlas 10:9 & N1467345916 & 128411.03 &  -6.06 & 0.35 & 39.93 &  0.51 & 144.2 &  9.50 &  0.72 & 26.9 &  11.32\\
Pan 19:18$^c$ & N1467345975 & 128897.87 &  -2.10 & 0.26 & 39.43 &  0.52 & 167.6 &  8.33 &  1.03 & 22.0 &  13.63\\
Prometheus 9:8 & N1467345975 & 128945.58 &  -5.02 & 0.71$^d$ & 46.53 &  1.02 & 138.3 &  9.02 &  0.93 & 33.1 & 195.57$^d$\\
Pan 22:21 & N1467346034 & 129541.69 &  -2.94 & 0.30 & 40.87 &  0.66 & 113.8 &  8.14 &  1.25 & 21.7 &  16.28\\
Pan 23:22 & N1467346034 & 129718.71 &  -3.83 & 0.63 & 41.52 &  1.87 &  96.5 &  8.16 &  1.07 & 21.4 &  14.38\\
Pandora 8:7 & N1467346034 & 129746.97 &  -4.70 & 0.64$^d$ & 41.67 &  0.70 & 123.0 &  9.08 &  1.13 & 34.2 & 142.49$^d$\\
Pan 24:23$^b$ & N1467346034 & 129880.87 &  -5.80 & 0.57 & 47.93 &  2.23 &  79.2 &  7.28 &  1.23 & 25.3 &   9.38\\
Pan 27:26 & N1467346093 & 130294.86 &  -4.57 & 0.44 & 43.49 &  1.09 &  24.6 &  7.74 &  1.30 & 21.8 &  12.67\\
Pan 29:28 & N1467346093 & 130523.00 &  -4.82 & 0.37 & 40.78 &  1.14 & 349.7 &  7.56 &  1.17 & 20.9 &   9.79\\
Pan 34:33 & N1467346152 & 130975.36 &  -2.23 & 0.47 & 40.16 &  1.31 & 259.1 &  7.60 &  0.80 & 19.4 &   9.64\\
Pan 35:34 & N1467346152 & 131050.25 &  -2.71 & 0.11 & 40.90 &  0.36 & 241.6 &  7.15 &  0.98 & 20.5 &  12.96\\
Pandora 9:8 & N1467346152 & 131101.74 &  -5.83 & 1.04$^d$ & 42.43 &  1.42 & 255.3 &  8.70 &  0.99 & 35.5 & 102.75$^d$\\
Pan 42:41 & N1467351539 & 131474.25 &  -4.56 & 0.90 & 50.58 &  4.48 & 118.4 &  6.30 &  1.25$^f$ &  6.3 &   9.04\\
Prometheus 12:11 & N1467346211 & 131590.30 &  -5.68 & 0.85$^d$ & 48.12 &  1.71 &  15.7 &  7.95 &  1.15 & 36.1 & 161.36$^d$\\
Prometheus 12:11 & N1467351539 & 131590.30 &  -3.30 & 0.92$^d$ & 44.10 &  1.75 & 324.2 &  6.96 &  1.10 & 10.3 & 147.91$^d$\\
Pan 55:54 & N1467351539 & 131974.51 &  -2.91 & 0.45 & 40.20 &  2.05 &  31.8 &  5.38 &  1.39$^f$ &  5.8 &  14.31\\
Pan 56:55 & N1467351539 & 132003.34 &  -3.96 & 0.37 & 45.77 &  1.83 &  52.8 &  4.58 &  1.14$^f$ &  7.2 &  25.55\\
Pan 58:57 & N1467351539 & 132058.02 &  -3.40 & 0.41 & 43.11 &  2.12 &  94.9 &  4.19 &  1.24$^f$ &  7.5 &  30.00\\
Pan 59:58 & N1467351539 & 132083.96 &  -3.54 & 0.54 & 43.90 &  3.26 & 115.9 &  4.71 &  1.39$^f$ &  6.6 &  24.41\\
Pan 60:59 & N1467351539 & 132109.04 &  -3.33 & 0.69 & 42.38 &  4.67 & 136.9 &  4.52 &  1.68$^f$ &  6.7 &  26.95\\
Pan 61:60 & N1467351539 & 132133.29 &  -2.74 & 0.38 & 38.99 &  2.10 & 158.0 &  4.46 &  1.61$^f$ &  6.5 &  22.42\\
Pan 62:61 & N1467351539 & 132156.76 &  -3.31 & 0.59 & 43.28 &  3.85 & 179.0 &  3.62 &  1.48$^f$ &  8.4 &  45.14\\
\hline
\end{tabular}
\end{center}
\begin{flushleft}
\vspace{-0.1in}
$^a$ The quality of the fit to this wave is affected by variations in background brightness (see Section~\ref{FitReal}).    
\\$^b$ The quality of the fit to this wave is affected by the nearby edge of the image (see Section~\ref{FitReal}).    
\\$^c$ The quality of the fit to this wave is affected by nearby density waves or other radial structure (see Section~\ref{FitReal}).
\\$^d$ This wave exhibits significant non-linearity  (``peakiness''), which affects the quality of the fits to $\Delta \rres$ and $A_L$.
\\$^e$ For this wave, $\xi_D$ was fit simultaneously with $A_L$, as described in Section~\ref{Amp}.
\\$^f$ At the location of this wave's maximum amplitude, the wavelength is less than 10 pixels, which may artificially lower the inferred $\xi_D$ (Section~\ref{ViscosityResults}).  
\end{flushleft}
\end{footnotesize}
\end{table}

\subsection{Surface Density \label{SurfDens}}

Our measurements of the background surface density (\Fig{s}~\ref{soirings_fitresults}a and~\ref{soirings_fitresults_alla}) are similar in magnitude to previous measurements made using stronger waves in \Voyit{} data \citep{Espo83,Lissauer85,NCP90,Rosen91a,Spilker04}, but show significantly less scatter.  We find a gentle linear trend of 0.0013 g/cm$^2$/km through the inner and mid-A~Ring (\Fig{}~\ref{soirings_fitresults_alla}), while analysis of waves in the outer A~Ring (which have too few wavecycles to be well characterized by our method) by other authors indicates that $\sigma_0$ decreases again exterior to the Encke Gap.  The goodness-of-fit for this linear trend (reduced $\chi^2 = 11$, with 29 degrees of freedom) suggests that the error estimates we quote for $\sigma_0$ are roughly 3 times too small, if indeed the linear trend is real.  On the other hand, it may well be that at least some of the deviation from the linear trend reflects real structure in the A~Ring.  

The weak waves (which are more likely to give well-behaved estimates of ring parameters -- see Section~\ref{SecondOrder}) in the unlit-side images (which have 3 to 4 times better resolution than the lit-side images -- see Table~\ref{ObserveInfo}) hew most closely to the linear fit, with a standard deviation of 2.1~g/cm$^2$.  Several of the significant outliers are relatively stronger waves, which may raise their own background surface density as very strong waves are known to do (Section~\ref{FirstOrder}), indicating that the linear trend may be a more accurate reflection of the unperturbed surface density.  

A correlation can be observed between the size of the error bar and the damping parameter $\xi_D$.  This is because a lower value of $\xi_D$ denotes a smaller number of wavecycles before the wave damps away; consequently, the wavelength dispersion is less accurately measured, driving up the uncertainty in $\sigma_0$.  

\subsection{Damping Parameter and Viscosity \label{ViscosityResults}}

\Fig{}~\ref{soirings_fitresults}b plots the damping parameter, giving a systematic survey of viscosity in the region between the Cassini Division and the Encke Gap.  \Fig{}~\ref{soirings_fitresults}c plots the viscosity as derived from \Eqn{}~\ref{DWViscosity}, showing a consistent increase over two orders of magnitude within the surveyed region.  Again we have agreement with previous measurements \citep{Espo83,LSC84,Shu85,Chakra89}, though the trend is much clearer in our data.  We note that \citeauthor{Espo83}'s three innermost data points, which are discordant with our data, come from waves --- Pandora~5:4, Prometheus~6:5, and Janus~4:3 --- that are now known to have highly irregular shapes (\Pone), and thus to be unsuitable for such analysis.  

Caution must be exercised in measuring the damping parameter $\xi_D$ for a wave with wavelengths not much larger than the image resolution.  Since a spiral density wave's wavelength decreases downstream (Section~\ref{WavenumberPhase}), the apparent amplitude can reach a local maximum because peaks and troughs are blurred as they become increasingly close together, not because of the actual physical damping process.  Since $\xi_D$ is obtained by locating the point of maximum amplitude (Section~\ref{LowFreq}), the wavelength at that location can be found by combining \Eqn{s}~\ref{Def_Xi}, \ref{Scripty_D}, \ref{DWWavenum}, and~\ref{Xi_max}:
\begin{equation}
k_{max} = 3^{-1/3} \left( \frac{3(m-1)}{2 \pi G \sigma_0 \rres} \right)^{1/2} \Omegares \xi_D
\end{equation}
\noindent This characteristic wavelength ($\lambda_{max} = 2 \pi / k_{max} $) is given for each wave in Table~\ref{WaveFitTable} as a ratio with the image resolution.  Most waves in our data set are well-resolved, with generally 20 or more pixels per wavecycle at the maximum amplitude point.  The waves in the lit-side images (Table~\ref{ObserveInfo}), however, are much more poorly resolved, with generally less than 10 pixels per wavecycle (keeping in mind also that the nominal image resolution is an upper limit; degradation can occur due to smearing, or imperfect pointing when taking radial scans).  These waves also yield much higher viscosities than any previous author has reported for the A~Ring.  Given that poor resolution would tend to boost the apparent viscosity, we do not plot these waves in \Fig{}~\ref{soirings_fitresults}b and~\ref{soirings_fitresults}c, though we do report the $\xi_D$ values in Table~\ref{WaveFitTable}.  

The rms velocity between ring particles can be derived from the viscosities using \Eqn{}~\ref{ViscosityRMSVelocity}, and the vertical scale height of the rings is simply $H \sim c/\Omega$ under the naive assumption that random velocities in the rings are isotropic.  However, self-gravity wakes \citep{JT66,Salo95}, which play a dominant role in driving viscosity in the A~Ring \citep{Daisaka01}, cause random velocities to be larger within the radial plane than in the vertical direction \citep{DI99}, thus depressing the vertical scale height implied by a given magnitude of rms velocity.  Hence, our "apparent thickness" should be interpreted as an upper limit for the A~Ring.  

This ``apparent ring thickness'' can also be read from \Fig{}~\ref{soirings_fitresults}c.  To allow viscosity and rms velocities to appear on the same plot, we used an average radial value to obtain $\Omega$, rather than using the actual orbital radius of each feature; this introduces errors of up to 4\%.  We also used an average optical depth $\tau = 0.5$.  

For the Cassini Division, however, where optical depths are too low for self-gravity wakes to be important, the implied vertical scale heights of 3.0~meters (Atlas~5:4) and 4.5~meters (Pan~7:6) are most likely correct as stated.\fn{As these tightly-wound waves have only $\sim 10$ pixels per wavecycle, their derived viscosities also may be anomalously high (see above).  In this case, the implication is that the Cassini Division's vertical scale height could be even lower than reported here.}  The apparent thicknesses in the inner A~Ring, consistently below 15~meters, also provide interesting constraints.  However, the mid-A Ring's trend upwards to 35~meters near the Encke Gap probably gives larger-than-real thicknesses.  

\citet{Daisaka01} parameterize ring viscosity from self-gravity wakes as observed in their simulations.  Using Saturn's mass $M_S = 5.69 \times 10^{26}$ kg and the density of ice $\rho = 0.9$~g/cm$^2$, their expression reduces to
\begin{equation}
\label{DaisakaPredict}
\nu(r) \simeq  26 \left( \frac{r}{122,000~\mathrm{km}} \right)^5 \frac{G^2 \sigma_0^2}{\Omega_0^3} .  
\end{equation}
\noindent This estimate, which implicitly includes a prediction of the abundance of self-gravity wakes, is also plotted in \Fig{}~\ref{soirings_fitresults}c.  Our data correspond fairly well, falling somewhat below the predicted viscosity in the inner A~Ring ($r < 127,000$~km), indicating that self-gravity wakes are somewhat less abundant there than \citeauthor{Daisaka01} predicted.  The reverse is true in the mid-A~Ring.  We note that the peak of the azimuthal brightness asymmetry, another indicator of self-gravity wakes, occurs at $r \sim 130,000$~km \citep{Dones93}.  

\begin{table}[!t]
\begin{footnotesize}
\begin{center}
\caption{Central pixel orbital radius for selected \Cassit{} images, from fitted density wave parameters in Table~\ref{WaveFitTable}, which assume the reverse-contrast regime (Section~\ref{ContrastRegime}).  For images containing more than one wave, the right-hand column shows the ratio between the uncertainty in $r_C$ and that predicted from the individual uncertainties in $\rres$.  \label{ImagePointTable}}
\vspace{0.1in}
\begin{tabular} { c r @{$\pm$} l c c }
\hline
\hline
Image & \multicolumn{2}{c}{$r_C$ (km)$^a$} & \# waves & $\sigma_i / \sigma_i'$ \\
\hline
N1467345208 & 118735.55 & 0.22 &  1 & \\
N1467345326 & 120489.23 & 0.11 &  1 & \\
N1467345621 & 124433.54 & 2.23 &  2 &  8.19\\
N1467345739 & 126012.76 & 0.32 &  2 &  1.60\\
N1467345798 & 126749.02 & 0.37 &  3 &  2.84\\
N1467345857 & 127488.91 & 0.27 &  1 & \\
N1467345916 & 128266.34 & 0.35 &  1 & \\
N1467345975 & 129009.84 & 0.26 &  1 & \\
N1467346034 & 129707.68 & 1.33 &  3 &  3.14\\
N1467346093 & 130398.98 & 0.18 &  2 &  0.44\\
N1467346152 & 131086.02 & 0.16 &  2 &  0.99\\
N1467351539 & 131976.81 & 0.52 &  8 &  1.08\\
\hline
\end{tabular}
\end{center}
\begin{flushleft}
\vspace{-0.1in}
$^a$ Orbital radius associated with the central pixel [511.5,511.5] of the image
\end{flushleft}
\end{footnotesize}
\end{table}

\subsection{Navigating Images}

Navigation data on the spacecraft's position and orientation, while excellent, are often accurate only to within tens of pixels for \Cassit{}'s Narrow Angle Camera (1 NAC pixel = 6 $\mu$rad).  Fiducial features appearing within an image are required for more accurate navigation, but truly circular features are rare in the A~Ring.  Here we use the density waves analyzed by our method as fiducial features to more accurately navigate the images in which they appear, by comparing the resonance locations determined by our method (i.e., where they are in the image) to the resonance locations that can be calculated from external parameters (i.e., where they should be in relation to Saturn and the rings).  In images from the highest-resolution unlit images (Table~\ref{ObserveInfo}), lines of constant orbital radius appear nearly straight, so that fine navigation is only needed (or possible) in the radial direction.  For these images, it is sufficient to identify the Saturn-centered orbital radius $r_C$ for the central pixel.  This is done in Table~\ref{ImagePointTable}.

Several images contain more than one density wave, providing a cross-check on the accuracy of our measurements.  For such images, the quoted $r_C$ is the average weighted by the individual uncertainties, and the right-hand column in Table~\ref{ImagePointTable} gives the re-normalization of the uncertainties as indicated by the agreement among the measurements.  The average renormalization indicates that we have underestimated the uncertainty of $\rres$ by a factor of 3 (we came to a similar conclusion regarding our error estimates for $\sigma_0$ in Section~\ref{SurfDens}).  

Excluded from this analysis are all waves due to first-order resonances of the larger moons, whose deviations from linear theory (Section~\ref{FirstOrder}) affect the quality of the fits for $\rres$.  

\section{Conclusions \label{Conclusions}}

We have provided a primer on how wavelet analysis can be employed to study density waves and other radial structure in Saturn's rings.  With this technique we have illuminated a number of ring features, as well as fitted the wave parameters of 32 density waves, most of them previously unobserved.  

Our results indicate a gently increasing trend of surface densities across the inner and mid-A~Ring.  We show viscous damping (and thus in-plane velocity dispersion) increases consistently from the Cassini Division to the Encke Gap, and place strict limits on the vertical thicknesses of the Cassini Division and the inner A~Ring.  We use the measured origin (resonance location) of each density wave to accurately navigate those images in which density waves appear that are suitable for our procedure.  

The final measurable parameter, the wave's amplitude, can be used to infer the mass of the perturbing moon, but only when the photometry is sufficiently well understood to relate observed brightness to ring optical depth and surface density.  This will likely be carried out in a later paper.  

The 32 density waves that are quantitatively analyzed in this paper represent only a fraction of the previously unobserved density waves that can be discerned in \Cassit{}~ISS data using wavelet analysis.  Some examples are shown here (see \Fig{s}~\ref{PanOLR}, \ref{SeparatingWaves1}, \ref{SeparatingWaves2}, and~\ref{PanWakes}), in order to illustrate the power of wavelet analysis to identify waves that are very weak or packed closely together.  We hope to undertake a more systematic survey of these in the future.  

\acknowledgements{We thank J.~Harrington, C.~Torrence and G.~Compo for helpful discussions.  We thank M.~W.~Evans for help with software development, and E.~Baker for technical assistance with the radial scans.  We thank J.-M.~Petit, J.~Cuzzi, and J.~Scargle for significant improvement to the manuscript.  We acknowledge funding by \Cassit{} and by NASA~PG\&G.}

\end{document}